\documentclass[journal]{IEEEtran}

\ifCLASSINFOpdf
\else
   \usepackage[dvips]{graphicx}
\fi
\usepackage{url}

\usepackage{amsmath, graphicx}
\usepackage{comment}
\usepackage{cite}
\usepackage{color}
\usepackage{enumerate}
\usepackage{array}
\usepackage{breqn}
\usepackage{amssymb}

\usepackage{multirow}
\usepackage{arydshln}
\begin{document}

\title{Interpretable Representation Learning for Speech and Audio Signals Based on Relevance Weighting}

\author{Purvi Agrawal, \IEEEmembership{Student member, IEEE}, and Sriram Ganapathy, \IEEEmembership{Senior Member, IEEE}
% \thanks{This paragraph of the first footnote will contain the date on which you submitted your paper for review. It will also contain support information, including sponsor and financial support acknowledgment. For example, ``This work was supported in part by the U.S. Department of Commerce under Grant BS123456.'' }
\thanks{Manuscript submitted on Feb. 24, 2020, revised on Jul. 29, 2020, accepted on Sep. 13, 2020.

This work was funded by grants from the Department of Atomic Energy (DAE/34/20/12/2018-BRNS/34088).

P. Agrawal and S. Ganapathy are with the Learning and Extraction of Acoustic Patterns (LEAP) lab, Department of Electrical Engineering, Indian Institute of Science, Bangalore, India, 560012. e-mail: \{purvia, sriramg\}@iisc.ac.in }}
%\markboth{Journal of \LaTeX\ Class Files, Vol. X, No. X, November 2019}
\maketitle
\begin{abstract}
The learning of interpretable representations from raw data presents significant challenges for time series data like speech. In this work, we propose a relevance weighting scheme that allows the interpretation of the speech representations during the forward propagation of the model itself. The relevance weighting is achieved using a sub-network approach that performs the task of feature selection. A relevance sub-network, applied on the output of first layer of a convolutional neural network model operating on raw speech signals, acts as an acoustic filterbank (FB) layer with relevance weighting. A similar relevance sub-network applied on the second convolutional layer performs modulation filterbank learning with relevance weighting. The full acoustic model consisting of relevance sub-networks, convolutional layers and feed-forward layers is trained for a speech recognition task on noisy and reverberant speech in the Aurora-4, CHiME-3 and VOiCES datasets. \textcolor{black}{The proposed representation learning framework is also applied for the task of sound classification in the UrbanSound8K dataset}. A detailed analysis of the relevance weights learned by the model reveals that the relevance weights capture information regarding the underlying \textcolor{black}{speech/audio content. In addition, speech recognition and sound classification experiments}  reveal that the incorporation of relevance weighting in the neural network architecture improves the performance significantly.
\end{abstract}

\begin{IEEEkeywords}
Deep Representation Learning, Relevance Modeling, Raw Waveform Processing, Modulation Filtering, Acoustic-phonetics, Automatic Speech Recognition, \textcolor{black}{Urban Sound Classification}.
\end{IEEEkeywords}

\IEEEpeerreviewmaketitle

\section{Introduction}
\label{sec:intro}
Representation learning is the branch of machine learning that comprises of the methods that allow the learning of meaningful representations from raw data. With the growing interest in deep learning, representation learning using deep neural networks has been actively pursued~\cite{bengio2013representation}. The modalities like text and image have shown important successes for representation learning methods (for example, using word2vec models \cite{mikolov2013efficient}). However, representation learning for speech data has been more challenging due to the temporal and intrinsic variabilities of the signals. Due to these challenges, the most common representation used in state-of-art speech recognition systems continues to be the mel spectrogram. In this paper, we explore representation learning for speech directly from the raw-waveform using a novel modeling approach.

The broad set of representation learning works in the speech domain can be categorized as supervised and unsupervised. In supervised setting, the main direction pursued has been the learning of the acoustic filterbank parameters  from raw waveforms \cite{doss2013, sainath2013, tuske2014acoustic}. Compared to the acoustic models using the mel spectrogram features, these models are fed with raw speech signal and the first layer of the acoustic model performs a time-frequency decomposition of the signal. The supervised objective function is either a detection or a classification task \cite{sainath2013, hoshen2015speech, sainath2015cldnn}. In many of these cases, the initialization of the acoustic filter bank parameters are done using gammatone filters that closely approximate the hearing periphery \cite{tuske2014acoustic,sainath2015cldnn}. However, in spite of many of these efforts, the filter learning from raw waveform has not yielded performance improvements over the mel spectrogram features \cite{sainath2015cldnn,doss2013}.

For unsupervised learning of acoustic representations, Sailor et al. \cite{sailor2016filterbank} investigated the use of  restricted Boltzmann machines while Agrawal et al. \cite{agrawal2019unsupervised} explored variational autoencoders. The wav2vec method proposed in \cite{schneider2019wav2vec} explored unsupervised pre-training for speech recognition by learning representations of raw audio. In our previous work, we had also explored modulation filter learning using an unsupervised objective function~\cite{purvi2019skipConn}.  The approach of self-supervised learning (similar to the word2vec models for text) has also been pursued by Pascual et. al~\cite{pascual2019pase}. A comparison of several neural network architectures for modulation filter learning was also explored in \cite{agrawal2018comparison}. 

%through two networks: encoder network and context network.  
In many of these previous approaches, the learned representations are assumed to capture distinctive speech features for phonetic discrimination. However, there has been limited attempts to explore the interpretability of the learned filterbanks. One exception is the SincNet filterbank \cite{ravanelli2018interpretable} where the authors showed that the filter learning is resilient to the presence of band-limited noise.  However, compared to vector representations of text which have shown to embed meaningful semantic properties, the interpretability of speech representations has been limited.
%Thus, it is unclear what information is captured in these learned representations. 

In this paper, we propose a relevance weighting mechanism that allows the interpretability of the learned representations in the forward propagation itself. 
%In the deep learning as whole, the role of interpretability can be partly through system modeling (eg. layer/kernel design) or can be through understanding the effect of model parameters on the task performance. For eg. after learning the acoustic filterbank from raw speech for speech recognition, it is still unknown about which frequency sub-bands contribute to a senone's (triphone target class) correct classification. Do some sub-bands contribute more than the other for a senone, where this distribution may change with change in target senone. Can we get this sub-band weighting information from the ASR model for a test speech signal? 
The term ``relevance weighting'' comes from the text domain, where  a static relevance weight is attached to each document, based on the relevance to the search term feature \cite{robertson1976relevance, robertson1997relevance}.   A similar application of visual attention in the image domain  uses spatial weighting to weight different parts of the image 
%(or re-weight the conv-layer feature map of a CNN). Its success is mainly due to the reasonable assumption that human vision does not tend to process a whole image in its entirety at once; instead only focuses on selective parts of the whole visual space, when and where as needed for a task
\cite{xu2015show}. In this work, the term relevance weighting  refers to the scheme of automatic weighting of the learned representations using a sub-network.

The proposed model consists of a two-step relevance weighting approach. The first step performs relevance weighting on the output of the first layer of convolutions. This convolutional layer learns a parametric acoustic filterbank from the raw waveform. 
% The acoustic filters are parametric cosine-modulated Gaussian filters whose parameters are learned within the acoustic model. The output of the layer generates a time-frequency representation similar to a spectrogram. 
% The output is fed to the relevance sub-network to obtain the relevance weights for the filterbank outputs. 
The relevance weighted filterbank representation is used as input to the second convolutional layer which performs modulation filtering. This layer repeats the operations of the first layer in a 2-D fashion. 
The kernels of the second convolutional layer are 2-D spectro-temporal modulation filters and the filtered representations are weighted using another relevance sub-network.
%The modulation filtered representations that come from the second layer are also weighted using a second layer of relevance sub-network.
%These weights  from the  second sub-network contain information regarding relevance of the modulation filters for the downstream task. 
The rest of the neural network architecture performs the task of acoustic modeling for speech recognition. All the model parameters including the acoustic layer parameters, acoustic filterbank relevance sub-network, modulation filters and modulation relevance sub-network are learned in a supervised learning paradigm.

The subsequent analysis of the relevance sub-network reveals that the weights from the network contain information regarding the phonetic content of the label. The relevance weights are also adaptive to the presence of noise in the data. 
% and the weights tend to favor the representations which have a higher signal to noise ratio (SNR). 
The automatic speech recognition (ASR) experiments are conducted on Aurora-4 (additive noise with channel artifact), CHiME-3 (additive noise with reverberation) and VOiCES (additive noise with reverberation) datasets. These experiments show that the learned representations from the proposed  framework 
%of filter learning with weighting 
provide considerable improvements in ASR results over the baseline methods.
\textcolor{black}{The proposed representation learning framework is further extended to the task of urban sound classification (USC) on the Urbansound8K dataset where the goal is to classify short audio snippets into $10$ urban sound classes. In this task, the proposed approach shows improved classification performance and the weight analysis reveals distinctive audio characteristics that are  captured by the relevance weighting scheme.}
%and other robust front-ends. 
%In addition, we analyze the relevance weights provided by the two sub-network modules (for acoustic sub-band weighting and modulation filtered representation weighting) in the trained deep model for phoneme representation analysis over noisy TIMIT dataset. 
%We also investigate the performance of the proposed framework in a semi-supervised setting where availability of labeled data is limited. The approach of weighting sub-bands and weighting modulation filtered representations as two-step process for the task of ASR and analysis on underlying phonetic content has been attempted for the first time to the best of our knowledge.

The rest of the paper is organized as follows. Section \ref{sec:related_work} details the motivation and  related work.
Section \ref{sec:rep_learning} describes the proposed two-step representation learning approach using relevance weighting. This is followed by interpretability analysis of the proposed representation learning in Section \ref{sec:analysis}. Section~\ref{sec:experiments} describes the ASR experiments with the various front-ends. The discussion of the model is given in Section~\ref{sec:discussion}. This is followed by a summary in  Section \ref{sec:summary}.
%followed by the results. We conclude the work with summary in Section \ref{sec:summary}.

\section{\textcolor{black}{Motivation and} Related work} {\label{sec:related_work}
\textcolor{black}{
One of the major motivation for the proposed modeling approach of relevance weighting is the evidence of gain enhancement mechanism in human sensory system. 
Both physiological and behavioral studies have suggested that stimulus-driven neural activity in the
sensory pathways can be modulated in amplitude with attention \cite{kauramaki2007selective}. The recordings of event-related
brain potentials indicate that such sensory gain control or amplification processes play an important role
in tasks that involve attention \cite{de2018auditory}. The combined event-related brain potential and neuroimaging experiments
provide strong evidence that attentional gain control operates at an early stage of sensory processing \cite{hillyard1998sensory}. These evidences support feature selection theories of attention and provide a basis for
distinguishing between separate mechanisms of attentional suppression (of unattended inputs) and attentional facilitation (of attended inputs).}

\textcolor{black}{
The second motivation comes from prior works on Mixture of Experts (MoE) models \cite{jacobs1991adaptive,jordan1994hierarchical}. For neural networks the work proposed by Shazeer et al. \cite{shazeer2017outrageously} explored multiple parallel neural networks followed by a gating network which performs a combination of the outputs from the networks. This MoE model showed significant promise for a language modeling task. The self attention module successfully inducted in transformer models \cite{vaswani2017attention} also incorporates information from multiple streams using a linear combination.} 

\textcolor{black}{
Inspired by these human and machine learning studies, the proposed relevance weighting attempts to model gain enhancement in a neural architecture.  The frequency selectivity observed in auditory system can be modeled as relevance weighting in the acoustic filter bank layer while the cortical layer gain enhancement is attempted by modulating the relevance weights of the modulation filtering layer. While attention \cite{vaswani2017attention} and MoE models \cite{shazeer2017outrageously}   attempt a combination of input streams that are fed to successive layers, the proposed relevance weighting merely performs a gain enhancement without a linear combination. }

The learning of acoustic filterbank from raw waveform has been actively pursued in the last few years. 
The work reported by Palaz et al. proposed the use of convolutional neural networks on raw waveform to estimate phoneme class conditional probabilities \cite{doss2013}. Similarly, Sainath et al. \cite{sainath2013} used power spectrum as input to convolutional neural network (CNN) to learn the filterbank features.  For learning interpretable representations, the authors in \cite{gaussian2017seki} used Gaussian functions instead of mel-scale filterbanks with input being power spectra. 
%Such parametric approach allows for reduced learnable parameters and enhances interpretability. 
The work proposed in \cite{zeghidour2018learning} initializes the convolutional filters with Gabor wavelets before proceeding to learn the filter parameters. Ravanelli et al. \cite{ravanelli2018interpretable} showed the use of sinc filters as parametric filters, with only the lower and higher cutoff frequencies learned from data. This work is extended in  \cite{pascual2019pase} to learn filterbank representations in a self-supervised framework.

The approach of modulation filter learning  (modulation filters process the time-frequency representation as a 2-D image and perform filtering along time (rate) and frequency (scale) dimensions) 
%for noise robustness has been well explored. Both the approaches of hand-crafted design of the modulation filters or learning filters from the data has been attempted. One of the earliest use of temporal modulations (rate) was the RASTA filtering approach \cite{hermanskyb}. The spectro-temporal modulation (rate-scale) filters for feature extraction, for example, Gabor filtering \cite{kleinschmidt2003localized, ezzat2007spectro}, have shown further improvements for ASR. A data-driven approach for parameter selection of Gabor filter set has been studied in \cite{kovacs2015}. 
using the linear discriminant analysis (LDA) has also been explored to learn the
 temporal modulation filters in a supervised manner \cite{vuurenLda1997, hung2006optimization}. There have also been attempts to learn modulation filters in an unsupervised manner \cite{agrawal2019jstsp, sailor2016unsupervised, purvi2017MF1D}}. 
 
 The performance of the learned representations from previous efforts has often been comparable to the mel-spectrogram features. However, most of the previous attempts do not allow the interpretation of the learned representations or their ability to capture relevant information. 
%  In addition, the filterbank learning has been mostly shown for speech recognition in clean conditions. 

%In all these approaches, the modulation filtering outputs (filtered spectrograms) are weighed equally, i.e. there is no weighting of one filter more over the other. In this work, we analyze the effect of weighting different modulation filtered representations over speech recognition task and also analyze the weighting on different phonemes. The weighting of sub-bands and weighting modulation filtered representations as two-step process for the task of ASR has been attempted for the first time to the best of our knowledge.

%\vspace{-1.3cm}
\section{Relevance Weighting Based Representation Learning}{\label{sec:rep_learning}}
The block schematic of the proposed relevance weighting based representation learning is shown in Figure~\ref{fig:block_diag}.

\subsection{Acoustic Filterbank Learning with Relevance Weighting} 
\begin{center}
    \begin{figure}[t]
        \centering
        \hspace{-1cm}
        \includegraphics[trim={0.5in 2.1in 8.5in 0.8in}, clip, scale=0.37]{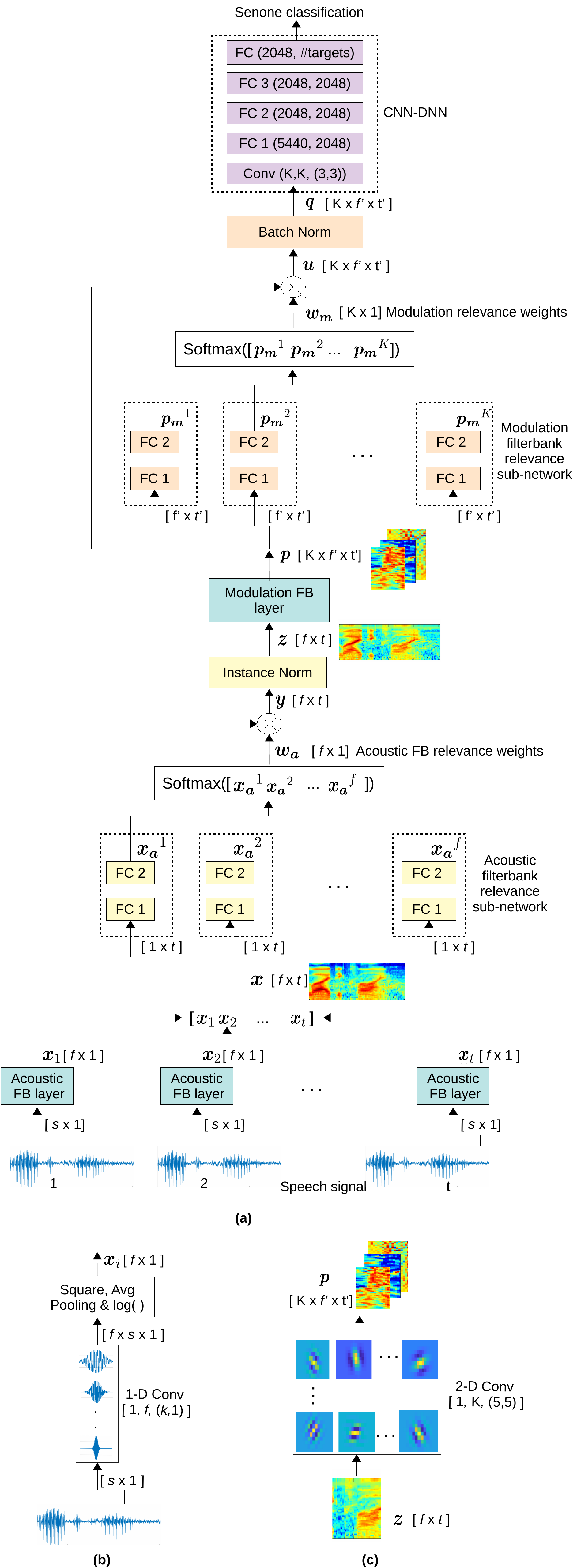}
        %\vspace{-0.9cm}
        \caption{(a) Block diagram of the proposed representation learning from raw waveform using relevance weighting approach. Here, FC denotes a fully connected layer and Conv denotes a convolution layer, (b) Expanded acoustic filterbank (FB) layer, (c) Expanded modulation FB layer.}
        \label{fig:block_diag}
        \vspace{-0.2cm}
    \end{figure}
\end{center}

 The first layer of the proposed  model performs acoustic filtering  from the raw waveforms using a convolutional layer.  The input to the neural network are raw samples windowed into $s$ samples per frame.
%  with a contextual window of $t$ frames. Each block of $s$ samples is refered to as a frame. 
This frame of raw audio samples are processed with a 1-D convolution  using $f$ kernels each of size $k$.  The kernels are modeled using a cosine-modulated Gaussian function \cite{agrawal2019unsupervised},
%\vspace{-0.05cm}
\begin{equation}
    {{g}}_i (n) = \cos{2\pi\mu_i n} \times \exp{(-{n^2}\mu_i^2/{2})}
    %\vspace{-0.05cm}
\end{equation}
where ${{g}}_i (n)$ is the $i$-th kernel ($i=1,..,f$), $\mu_i$ is the center frequency (mean parameter) of the $i$th kernel.
% and variance of the Gaussian function is tied to the mean as $\sigma_i = 1/\mu_i$.
% The number of filter taps is denoted as $k$.  
The parametric approach to filterbank (FB) learning generates filters with a smooth frequency response. The Gaussian window has interesting properties of being smooth in time and frequency domain. \textcolor{black}{In order to preserve the positive range of frequency values for $\mu$, we choose the $\mu$ to be the sigmoid of a real number which is further scaled to half the sampling frequency of the signal.}
% The location of the Gaussian kernel in the frequency domain (center frequency) is determined by the mean parameter $\mu_i$. 
The mean parameters are updated in a supervised manner for each dataset. 
%More discussion on initialization of filterbank and transfer learning is given in Section \ref{sec:discussion}.
The convolution with the cosine-modulated Gaussian filters generates $f$ acoustic feature maps. These outputs are squared, average pooled within each frame and log transformed. This generates $\boldsymbol{x}$ as $f$ dimensional features for each of the $t$ contextual frames, as shown in Figure \ref{fig:block_diag}. The matrix $\boldsymbol{x}$ is interpreted as the ``learned'' time-frequency representation.
% (spectrogram). We refer to the first layer as the acoustic filterbank (FB) layer and the output of each Gaussian kernel as an acoustic feature map.
%sized frame level feature vector. We then shift the window (101 times to keep splicing of +/- 50) around the center frame in raw waveform by a hop size of 10ms and repeat this convolution to produce time-frequency patches of size $80 \times 101$. The block diagram is shown in Figure \ref{fig:block_diag} where input is tensor $\boldsymbol{x}$ consisting of enframed raw signal and output of the acoustic filterbank layer is tensor $\boldsymbol{X}$ of size [B, 1, 80, 101] consisting of 2-D time-frequency (Spec) patches.
\begin{figure}[t]
    % \centering
    \includegraphics[trim={0.6in 0 0.65in 0.2in}, clip, scale=0.34]{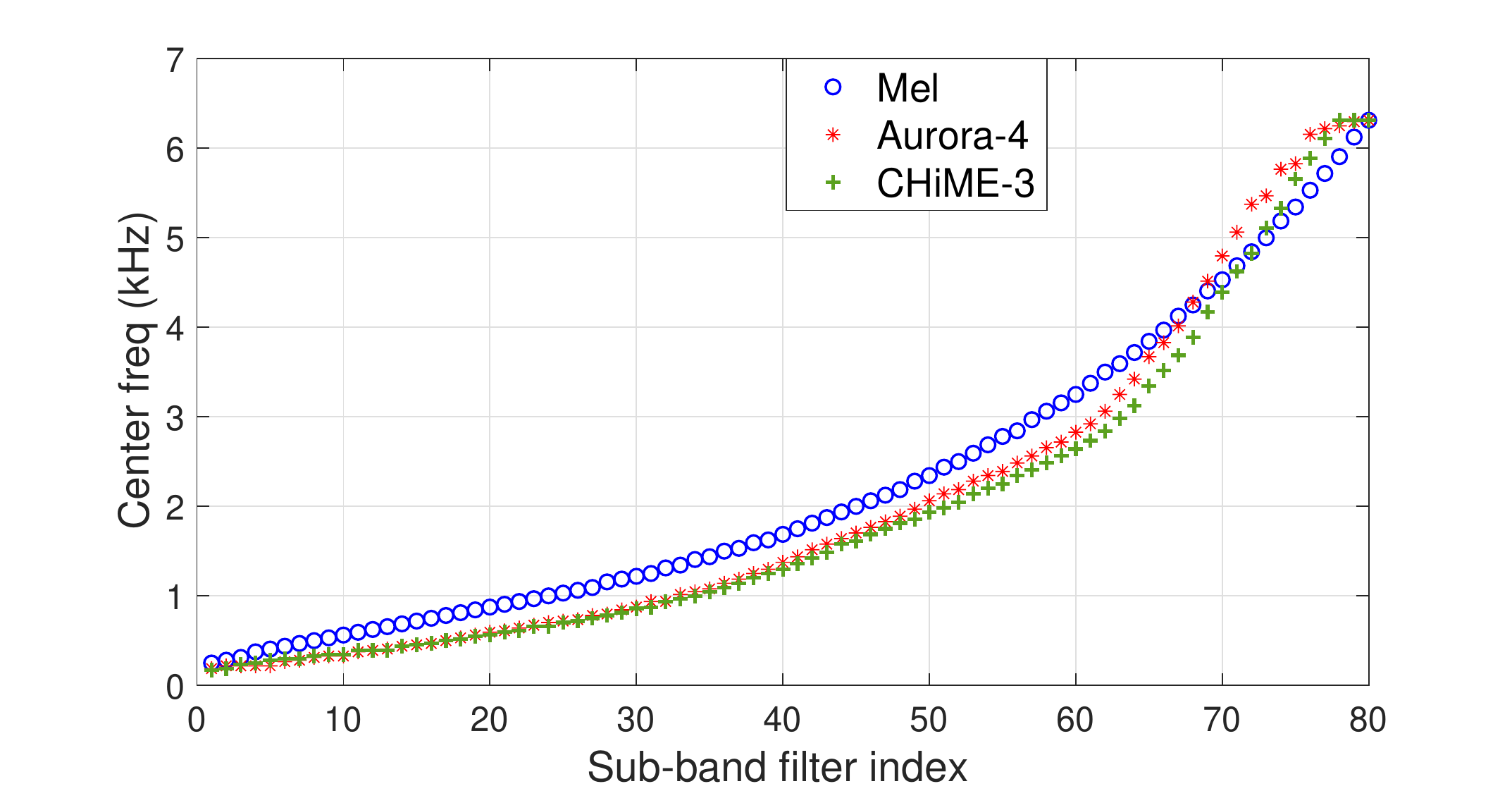}
    \hspace{-0.3cm}
    \includegraphics[trim={0.37in 1in 0.28in 0.1in}, clip, scale=0.4]{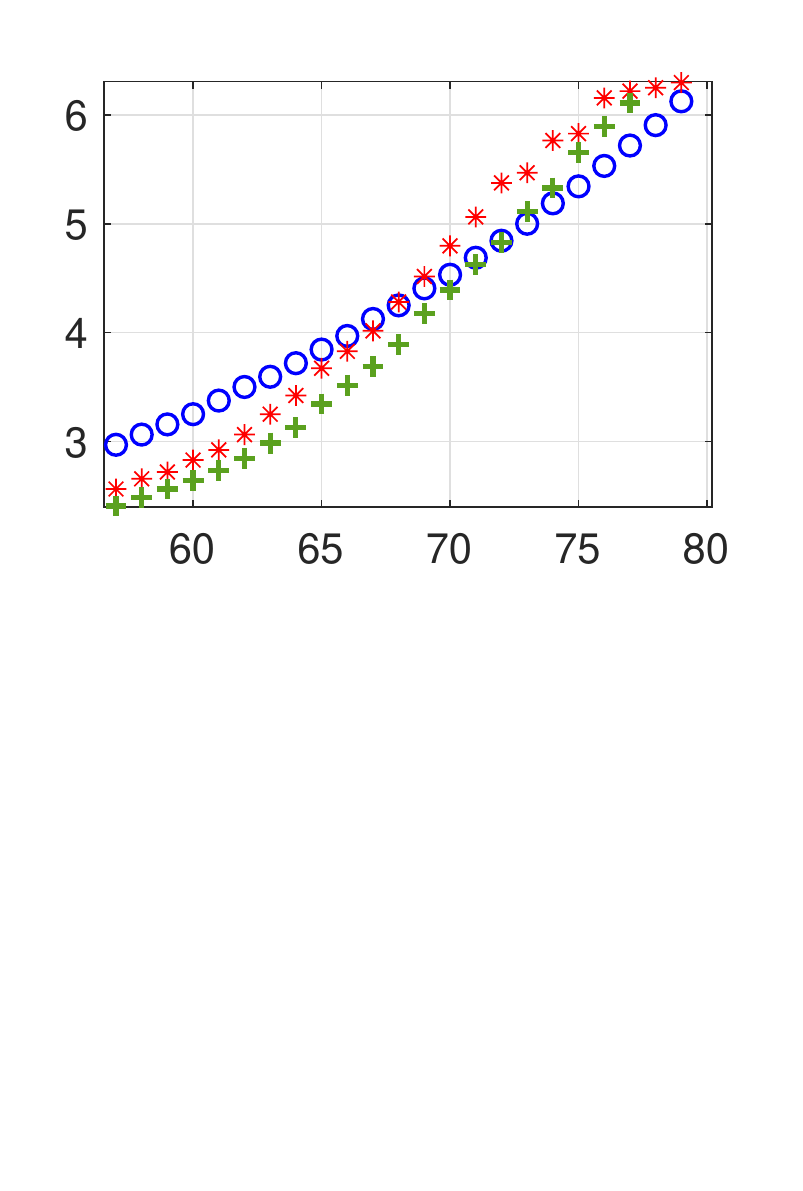}
    %\vspace{-0.2cm}
    \caption{Left Fig: Comparison of center frequency of acoustic filterbank learned using the proposed approach for Aurora-4 and CHiME-3 datasets, with center frequencies of mel filterbank. The filter indices are ordered in their increasing order of center frequency; \textcolor{black}{Right Fig: Zoomed portion of the left figure for sub-bands $55-80$. }}
    \label{fig:center_freq_acoustic}
     \vspace{-0.15cm}
\end{figure}
\vspace{-0.6cm}
\subsubsection{Relevance weighting} {\label{subsec:self_relevance weighting}}
The relevance weighting paradigm for acoustic FB layer is implemented using a relevance sub-network. This network is fed with the acoustic feature map of dimension $1 \times t$ for each of the $f$ Gaussian kernel (each row of the  time-frequency representation $\boldsymbol{x}$). A two layer deep neural network (DNN) with a $t$ dimensional input layer and a scalar output realizes the relevance sub-network. This operation is repeated for all the $f$ acoustic feature maps to generate $\boldsymbol{x_a}$ as a $f$ dimensional vector with weights corresponding to each kernel. The relevance weights $\boldsymbol{w_a}$ are generated using the softmax function as,
\begin{equation}
    w_a^i = \frac{e^{x_a^i}}{\sum_j e^{x_a^j}}; ~~\text{where}~ i = 1, 2, ..., f. 
\end{equation}
The weights $\boldsymbol{w_a}$ are multiplied with each of the acoustic feature map (rows of $\boldsymbol{x}$) to obtain the relevance weighted time-frequency  representation $\boldsymbol{y}$.
The relevance weights in the proposed framework are different from typical attention mechanism~\cite{zhang2017attention}. In {\textcolor{black}{the}} proposed framework, relevance weighting is applied over the representation as soft feature selection weights without a linear combination as done in attention models~\cite{zhang2017attention}. 

We also smooth the first stage outputs (weighted spectrogram $\boldsymbol{y}$ of length $t$ time frames) using a normalization method inspired by instance norm principle~\cite{rumelhart1986learning, ulyanov2016instance}. Let $y_{j,i}$ denote the relevance weighted time-frequency representation for frame $j$ ($j=1,..,t$) of kernel $i$ ($i=1,..,f$). The soft weighted output $z_{j,i}$ is given as,
%\vspace{-0.15cm}
\begin{equation}\label{eq:soft}
% \vspace{-0.15cm}
    z_{j,i} = \frac {y_{j,i} - m_{i}} {\sqrt{\sigma ^2 _i + c}} 
    %\vspace{-0.15cm}
\end{equation}
% \vspace{-0.15cm}
where $m_i$ is the sample mean of $y_{j,i}$ computed over $j$ and $\sigma _i$ is the sample std. dev. of $y_{j,i}$ computed over $j$. The constant $c$ acts as a relevance factor and is chosen as $10^{-4}$. 
% a hyper-parameter in the model. 
%When the relevance weight for sub-band $i$ is high, the std. dev. $\sigma _i$ is also high compared to $c$ and thus the soft weighted output $z_{j,i}$ has a unit variance over $j$. When the relevance weight for sub-band $i$ is low, the value of $\sigma _i$ is also relatively less compared to $c$ and this makes the variance of $z _{j,i}$ lower than $1$. 
% Thus, Eq.~\ref{eq:soft} modulates the relevance weighting mechanism and 
The output of relevance weighting ($\boldsymbol{z}$) is propagated to the subsequent layers for the acoustic modeling. 

In our experiments, we use $t=101$ whose center frame is the senone target for the acoustic model. We also use $f=80$ kernels each of length $k=129$. This value of $k$ corresponds to $8$ ms in time for a $16$ kHz sampled signal which has been found to be sufficient to capture temporal variations of a speech signal \cite{lewicki2002efficient}. The value of $s$ is $400$ corresponding to $25$ms window length and the frames are shifted every $10$ms. 
% Thus, the input to the acoustic filter bank layer with $t=101$ yields about $1$s context  of audio segment. 
{\textcolor{black}{The instance norm is not applied globally but rather applied on the given input patch of  $t=101$ time frames (shown in Fig. \ref{fig:block_diag}). This corresponds to $1$-second of length approximately.}} In our experiments, we also find that, after the normalization layer, the number of frames $t$ can be pruned to the center $21$ frames for the acoustic model training without loss in performance. This has significant computational benefits and the pruning is performed to keep only the $21$ frames around the center frame. 
% We empirically choose the value of relevance factor $c=10^{-4}$ in our work. 

Figure \ref{fig:center_freq_acoustic} shows the center frequency ($\mu_i$ values sorted in ascending order) of the acoustic filters obtained using multi-condition Aurora-4 (noisy) and CHiME-3 (noise and reverberations) datasets (details of the datasets are given in Section \ref{sec:experiments}) and this is compared with the center frequency of the conventional mel filterbank \cite{mfccdavis}. As can be observed, the proposed filterbank allocates more filters in lower frequencies compared to the mel filterbank. \textcolor{black}{The CHiME-3 data contains reverberation artifacts which resulted in lower center frequencies compared to the noisy Aurora-4 data center frequency values.}

\begin{figure*}[t]
    \centering
    %\vspace{-0.1cm}
    \includegraphics[trim={1cm 0cm 0cm 0.9cm}, clip, scale=0.425]{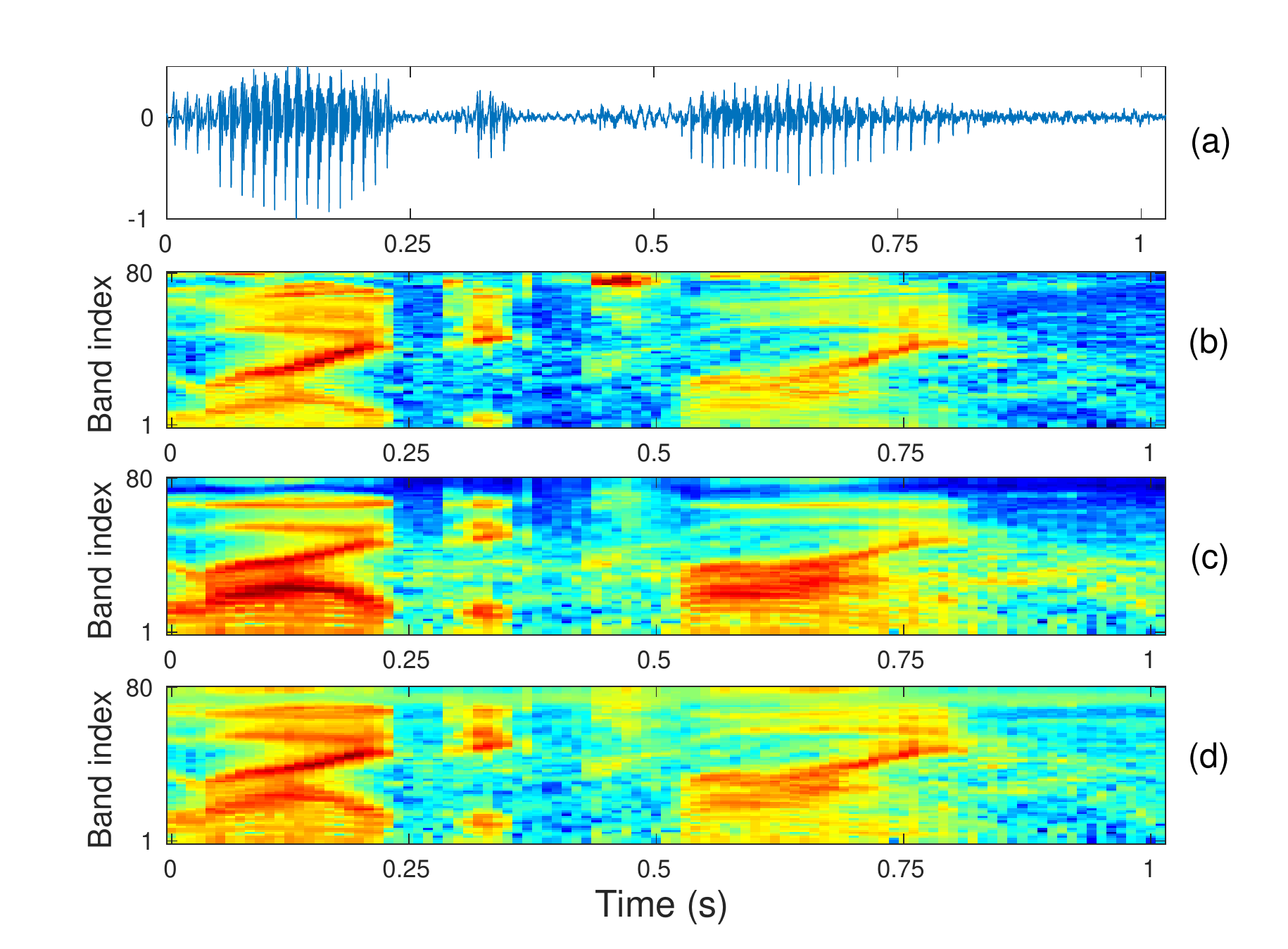}
    \includegraphics[trim={1cm 0.2cm 0cm 0.5cm}, clip, scale=0.4]{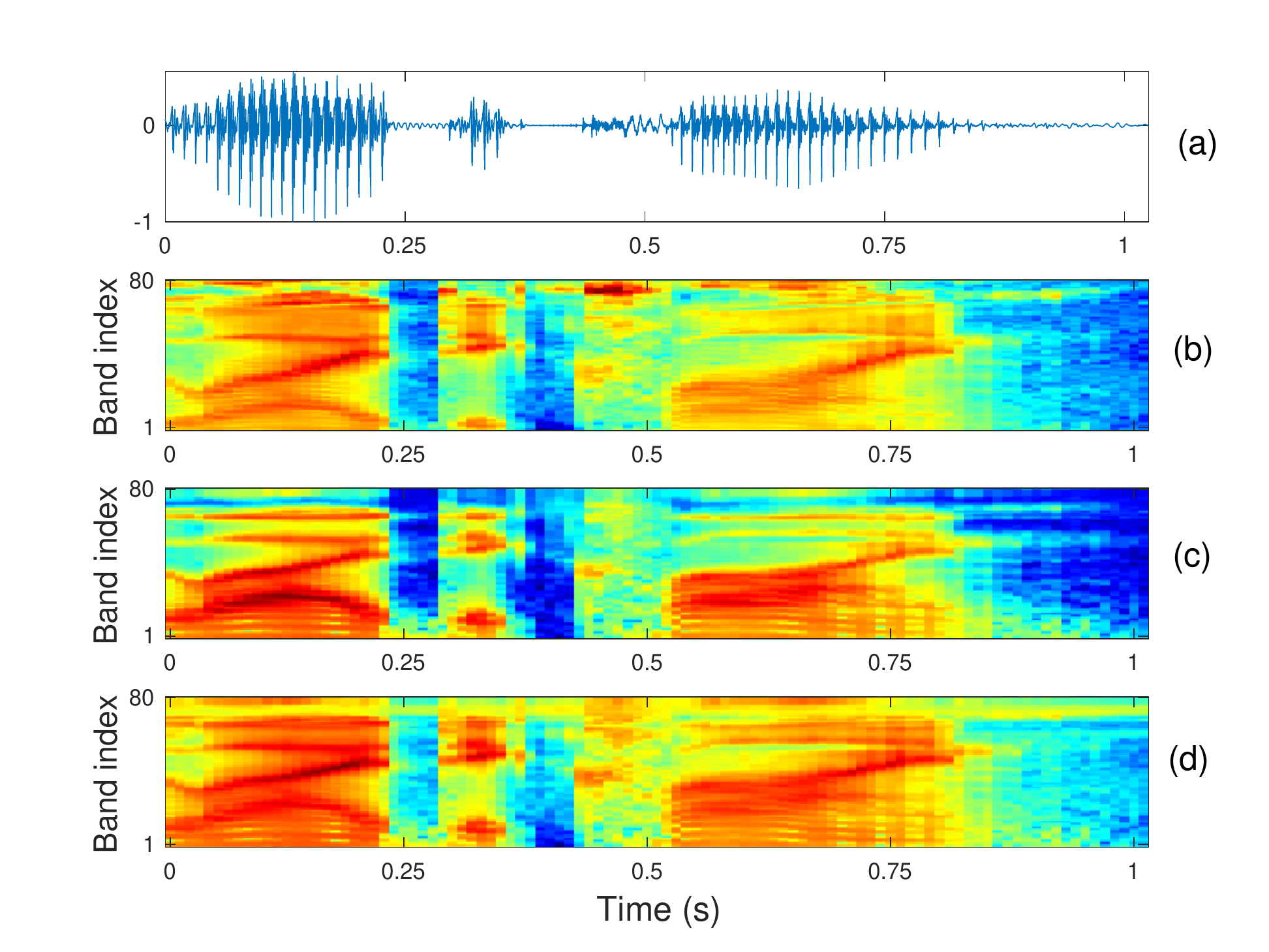}
    % \vspace{5cm}
    \includegraphics[trim={0cm 0.2cm 0cm 1.9cm}, clip, scale=0.338]{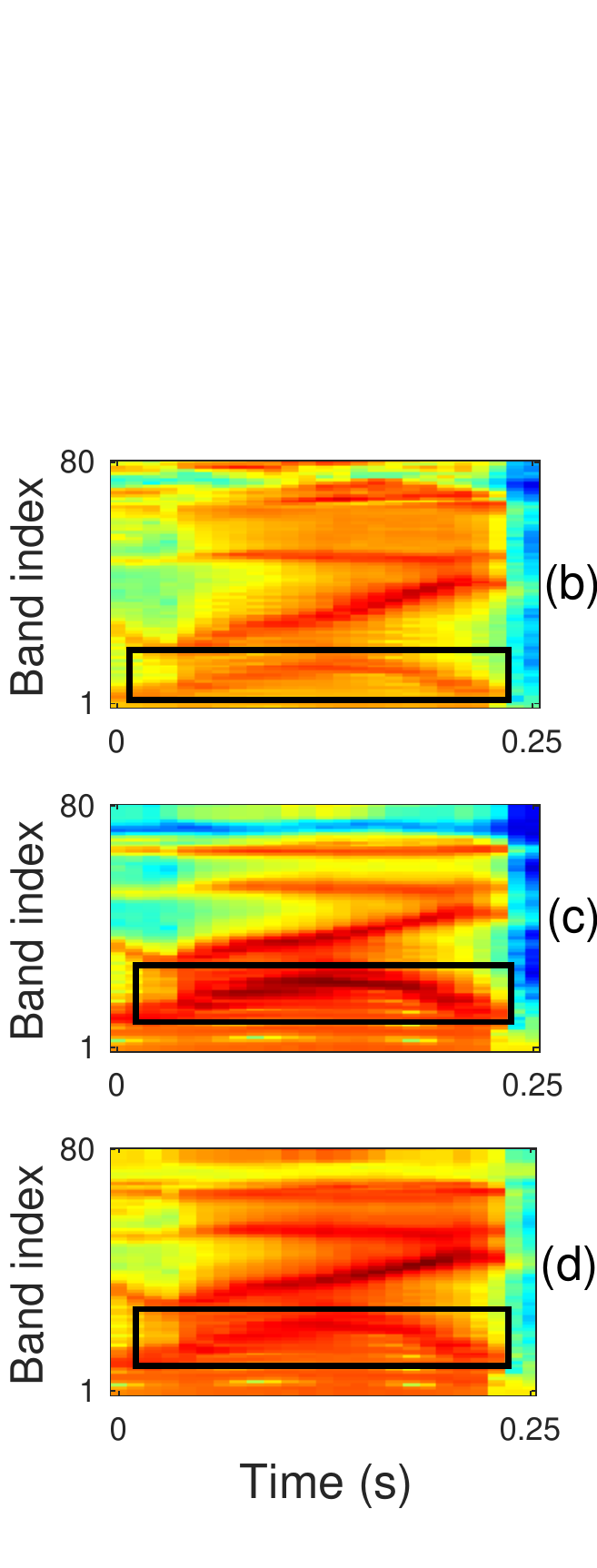}
    \vspace{-0.2cm}
    \caption{Left Fig: (a) Speech signal from Aurora-4 dataset with airport noise, (b) mel spectrogram representation (c) acoustic filterbank representation ($\boldsymbol{x}$ in Figure \ref{fig:block_diag}) (d) acoustic filterbank representation with soft relevance weighting ($\boldsymbol{z}$ in Figure \ref{fig:block_diag}), {\textcolor{black}{Middle Fig: Corresponding representations for a clean speech signal, Right Fig: Zoomed formant shift highlighted in rectangular box.}}}
    \label{fig:sig_spec} 
    \vspace{-0.35cm}
\end{figure*}

The soft relevance weighted time-frequency representation $\boldsymbol{z}$ obtained from the proposed approach is shown in Figure \ref{fig:sig_spec}(d) for an utterance with airport noise from Aurora-4 dataset (the waveform is plotted in Figure \ref{fig:sig_spec}(a)). The corresponding mel spectrogram (without relevance weighting) is plotted in Figure \ref{fig:sig_spec}(b). The time-frequency representation obtained through the learned filterbank (without relevance weighting) is plotted in Figure \ref{fig:sig_spec}(c). It can be observed that,  in the learned filterbank representations (Figure \ref{fig:sig_spec}(c) and (d)), the formant frequencies are shifted upwards because of the increased number of filters in the lower frequency region. Also, the relevance weighting modifies the representations to preserve only the important details of the spectrogram.

\subsection{Modulation Filterbank Learning with Relevance Weighting}
The representation $\boldsymbol{z}$ from acoustic filterbank layer is fed to the second convolutional layer which is interpreted as modulation filtering layer (shown in Figure \ref{fig:block_diag}). Specifically, the modulation filter is characterized using the rate (temporal variations measured in Hz) and scale (spectral variations measured in cyc. per octave) dimensions. 
\textcolor{black}{
The first layer (acoustic FB)  generates time-frequency representations (2-D representations) as the output ($\boldsymbol{z}$ in Fig. 1). These 2-D representations are indexed by time-frame on x-direction and sub-band index in the y-direction. The sub-band indices are ordered in increasing value of center frequency which is required for modulation filtering in the second layer.  The next stage of 2-D modulation filtering (rate-scale) is applied on 2-D time-frequency representation. The column of the time-frequency representation from the first layer constitutes a warped sampling of the spectrum from $0-8$ kHz with center frequencies shown in Figure 2. Hence, applying a filter along the y-direction of this 2-D time frequency representation constitutes scale filtering and the application of the filter along the x-direction constitutes rate filtering. Note that the sampling in the y-direction is non-linear in frequency.}\\

% The kernels of this convolutional layer are interpreted as 2-D spectro-temporal modulation filters, learning the rate-scale characteristics from the data. 
In this work, we explore parametric and non-parametric approaches to modulation filter learning.
% The modulation filtered representations are also selectively weighted using a second relevance sub-network as shown in Figure~\ref{fig:block_diag}. 
In the parametric approach, the modulation filterbank (kernels of the modulation convolution layer) is designed as 2-D cosine-modulated Gaussian filters. Here, the \textit{i}th 2-D filter $\boldsymbol{g}_i$ with rate frequency $\mu_{r_i}$ and scale frequency $\mu_{s_i}$ is designed as:
\begin{equation}{\label{eq:2D_cos_gauss}}
    {{g}}_i (a,b) = \cos{2\pi(\mu_{r_i} a \pm \mu_{s_i} b)} \times \exp{[(-{a^2}) + (-{b^2})]}
\end{equation}
with $a$ sampled at $100$ Hz (corresponding to $10$ ms hop in the time-frequency representation), $b$ sampled at $24$ cyc. per octave, and $i=1, ..., K$ for $K$ modulation filters. The $\pm$ sign in the cosine term is used to incorporate upward and downward moving patterns in the input time-frequency representation. The cosine frequencies $\mu_{r_i}$ and $\mu_{s_i}$ are interpreted as rate and scale center frequencies in 2-D frequency response of the modulation filter. The means are the learnable parameters of the 2-D kernels and these are learned jointly with rest of the network parameters. We learn $K=40$ modulation feature maps using kernels of $5 \times 5$ filter tap size.

\begin{figure}[t!]
    \centering
    \includegraphics[trim={0 0 0 0}, clip, scale=0.42]{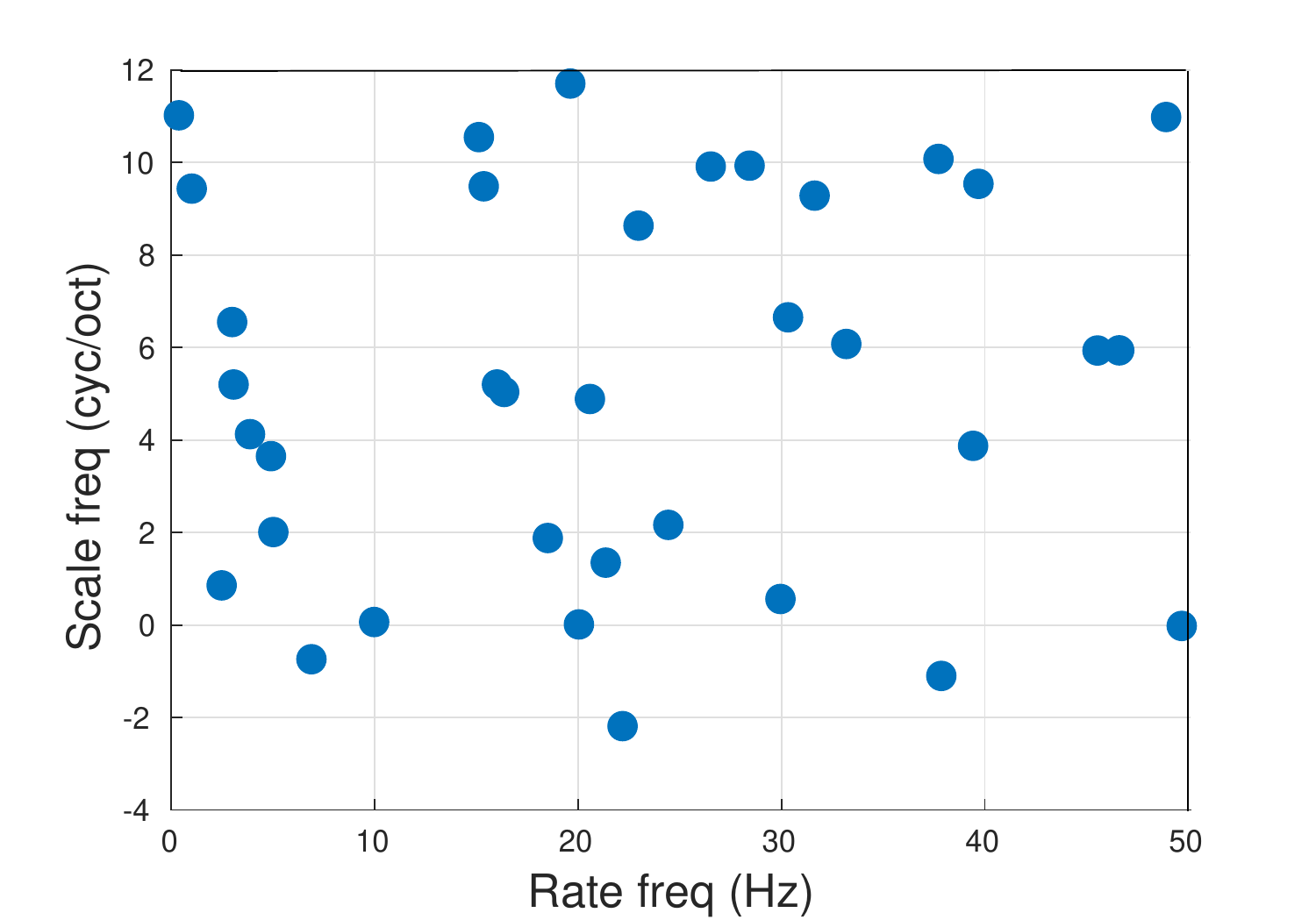}
    \vspace{-0.1cm}
    \caption{Center frequency of the 2-D parametric modulation filters learned using the proposed approach for Aurora-4 dataset.}
    \label{fig:mod_filters_fc}
    \vspace{-0.3cm}
\end{figure}

The learned center frequencies $\mu_{r}$ and $\mu_{s}$ (rate-scale) of the modulation filters are shown in Figure \ref{fig:mod_filters_fc}, from a model trained using Aurora-4 database.
% In the plot, the negative (positive) rate frequency in x-axis denotes upward (downward) moving direction of the corresponding harmonic ripple in the 2-D impulse response, modulated at corresponding rate-scale frequency.
The rate frequency values span upto $50$ Hz, while the scale frequency can span from $0$ to $12$ cycles/octave. It can be observed from the plot that the learned filters span the rate-scale space with more density till around $35$ Hz rate frequency. The learned center frequencies span low scale regions with most of the filters in the $[-2,6]$ cyc. per octave range. 
%The scale frequency can also take negative values, owing to the negative orientation (downward pattern) of the input time-frequency representation. 
We begin with randomly spaced grid of rate-scale values (center frequency) as initialization and let the network update the center frequency values. 
% The variances are kept fixed in the current setup for analysis.
% The center frequencies are symmetric across positive and negative rate, owing to the design of the 2-D modulation filters as described in Equation \ref{eq:2D_cos_gauss}.
\begin{figure*}[t]
    \centering
    \hspace{-3.5cm}
    \begin{minipage}{.1\textwidth}
    \includegraphics[trim={0 0 0 0}, clip, scale=0.28]{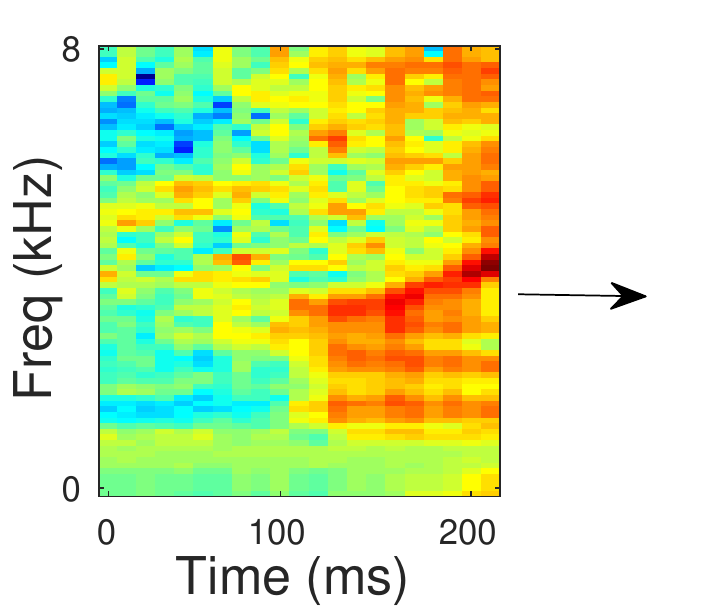}
    \end{minipage}%
    \hspace{0.15cm}
    \begin{minipage}{0.7\textwidth}
    \includegraphics[trim={0 0 0 0}, clip, scale=0.42]{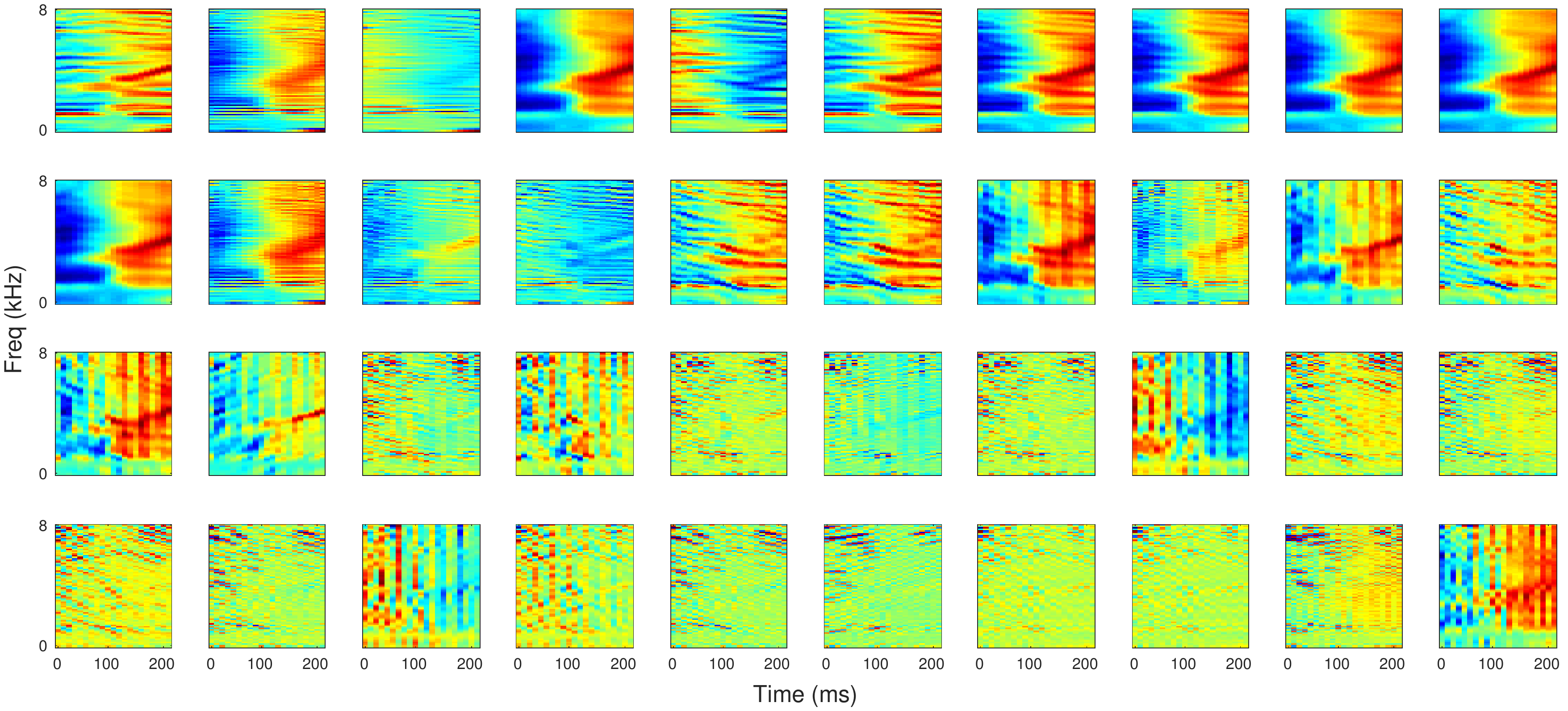}
    \end{minipage}
    %\vspace{-0.2cm}
    \caption{Plot of modulation feature maps ($\boldsymbol{q}$ in Figure \ref{fig:block_diag}) for an input patch (shown on the left, corresponding to $\boldsymbol{z}$ in Figure \ref{fig:block_diag}) for an utterance from Aurora-4 dataset - airport noise (feature maps plotted in order of increasing rate frequency). }
    \label{fig:mod_featureMap}
     \vspace{-0.3cm}
\end{figure*}
% The modulation filtering layer generates $K$ modulation feature maps, corresponding to $K$ modulation filters $\boldsymbol{w}_K$. 
The modulation feature maps $\boldsymbol{p}$ are pooled, leading to feature maps of size $f' \times t'$. These are weighted using a second relevance weighting sub-network (referred to as the modulation filter relevance sub-network in Figure \ref{fig:block_diag}). The input to the modulation relevance weighting sub-network is a modulation feature map of size  $f' \times t'$ and the model generates a scalar value for each of the $K$ feature maps.  Let $\boldsymbol{p_m}$ denote the $K$ dimensional vector from the output of modulation relevance sub-network. Similar to the acoustic relevance network, a softmax function is applied to generate modulation relevance weights $\boldsymbol{w_m}$,
%\vspace{-0.2cm}
 \begin{equation}
     w_m^i = \frac{e^{p_m^i}}{\sum_j e^{p_m^j}}; ~\text{where}~ i = 1, 2, ..., K. 
     %\vspace{-0.2cm}
 \end{equation}
The weights are multiplied with the representation $\boldsymbol{p}$ to obtain relevance weighted modulation feature maps $\boldsymbol{q}$. This weighting performs the adaptive selection of different modulation feature map representations (with different rate-scale characteristics). The resultant weighted representation $\boldsymbol{q}$ is fed to the batch normalization layer \cite{batch2015norm}.
% The training data statistics of batch norm, including affine parameters, 
%(i.e. gamma, beta) 
% are used in the test phase. 
%Here, the presence of relevance factor in the denominator of normalization modulates the weighting mechanism. 
The value of the normalization factor for batch norm is also chosen to be $10^{-4}$ empirically. 
% We use $K=40$ modulation filters. 
Following the acoustic filterbank layer and the modulation filtering layer (including the relevance sub-networks), the acoustic model consists of series of  CNN and  deep feed forward layers. The configuration details of different model parameters are given in Figure~\ref{fig:block_diag}. The entire model is trained using the cross entropy loss with Adam optimizer \cite{kingma2014adam}. 
%Hence, the resultant normalized representation $\boldsymbol{u}$ is then fed to the CNN-DNN architecture for the task of speech recognition. 

Figure \ref{fig:mod_featureMap} shows the obtained feature maps after the second stage of modulation filtering. The input to this layer is shown on left ($\boldsymbol{z}$ in Figure \ref{fig:block_diag}) and the obtained weighted $K=40$ feature maps ($\boldsymbol{q}$ in Figure \ref{fig:block_diag}) are plotted in the right. As can be observed, the modulation filtering layer filters the input patch using rate-scale filters with varying center frequencies. 
% The second stage outlined above is motivated by multi-stream feature framework for ASR \cite{mesgarani2006discrimination, nemala2013multistream}. The temporal and frequency patterns in spectrogram (called modulations) characterize several important cues associated with different sound percepts, such as slow rate commensurate with the syllable rate in speech, intermediate and fast rates capture segmental transitions like onsets and offsets. Similarly, slow/broad spectral modulations capture primarily the overall spectral profile and formants, while fast/narrow modulation scales reflect spectral details such as harmonics and sub-harmonic structure of the spectrum \cite{nemala2013multistream}.

The proposed two stage processing is loosely modeled based on our understanding of the human auditory system, where the cochlea performs acoustic frequency analysis while early cortical processing performs modulation filtering \cite{mesgarani2006discrimination}. The relevance weighting mechanism attempts to model the feature selection/weighting inherently present in the auditory system (based on the relative importance of the representation for the downstream task). 

\begin{figure}[t]
    \centering
    \includegraphics[trim={1in 0.1in 0.68in 0.1in}, clip, width=\linewidth]{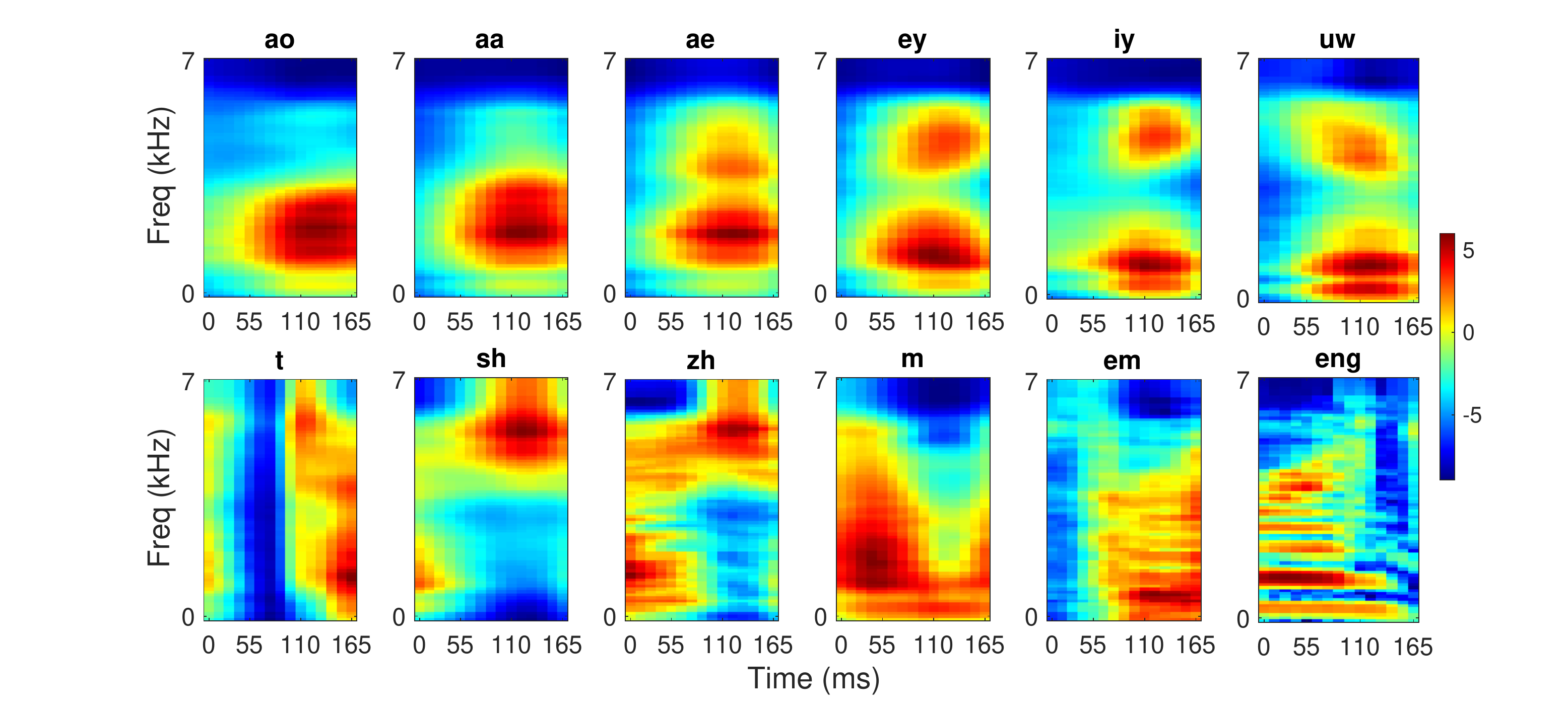}
    \vspace{-0.5cm}
    \caption{Average time-frequency representation learned by the model for vowel phonemes (top row) and consonant phonemes (bottom row) {\textcolor{black}{from the clean TIMIT test set.}}}
    \label{fig:avg_phoneme_spec}
    \vspace{-0.2cm}
\end{figure}
\section{Interpretability of the Representations}{\label{sec:analysis}}
We analyze the representations and the relevance weights (outputs of acoustic relevance  sub-network and modulation relevance  sub-network) learned by the proposed model. The model architecture, shown in Figure~\ref{fig:block_diag}, is trained using the Aurora-4 dataset. This model is tested using the TIMIT dataset (without any retraining). \textcolor{black}{We use clean TIMIT as well as TIMIT dataset corrupted with noise}  \cite{timit1993darpa}. The TIMIT dataset is hand labelled for phonemes, which allows the interpretation of the model representations based on the phoneme identity. For the analysis, the noisy TIMIT  test set consisting of $1344$ utterances is corrupted with $4$ different noise conditions, namely babble, exhall, restaurant and subway at $2$ different SNR levels, $0$ dB and $20$ dB SNR. 
%The trained model from Aurora-4 dataset (details of the dataset is given in Section \ref{sec:experiments}) is used for the analysis by forward-pass of the Noisy TIMIT data through the trained model and analysing the intermediate representations as well as sub-network relevance weights for phonemes. 

The phonemes are analyzed in two groups.  A group of vowel phonemes \{/ao/, /aa/, /ae/, /ey/, /iy/, /uw/\} and a group of  consonants from $3$ categories: plosives \{/t/\}, fricatives \{/sh/, /zh/\}, and nasals \{/m/, /em/, /eng/\}. The vowel sounds are also organized from back to front with regard to the place of articulation~\cite{mesgarani2008phoneme}. 
%In the Figure \ref{fig:avg_phoneme_spec}, the vowels in top row are organized according to their articulatory configuration along the front/back axes \cite{mesgarani2008phoneme}. 
\begin{figure*}[t]
    \centering
    %\vspace{-0.2cm}
    \includegraphics[trim={5.1cm 0 4.5cm 0cm}, clip, width=\linewidth]{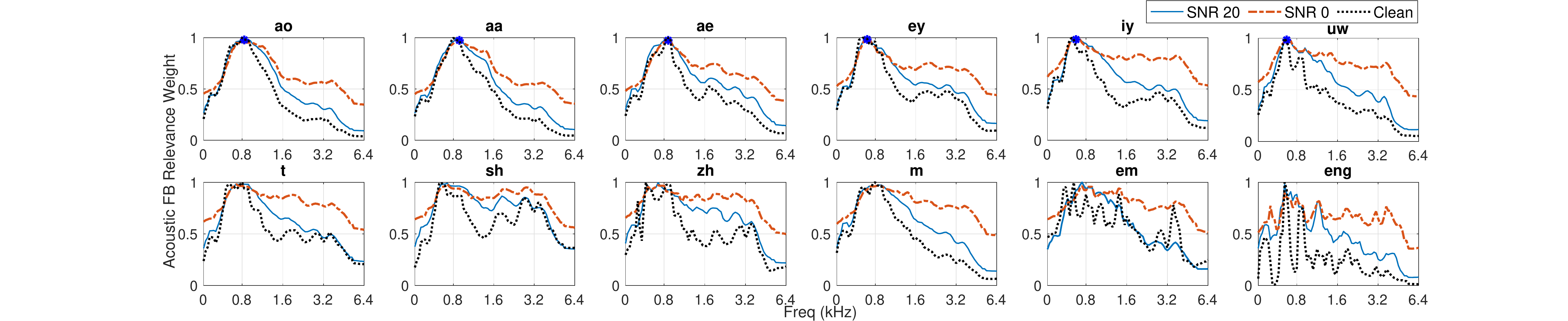}
    \vspace{-0.6cm}
    \caption{The normalized acoustic FB relevance weight profile for each phoneme: vowels in top row and consonants in bottom row, computed using the relevance weights for {\textcolor{black}{clean (black dotted) and noisy TIMIT files with SNR $20$ dB (black solid) and SNR $0$ dB {\textcolor{black}{(red dot-dashed)}}}}, respectively.}
    \label{fig:best_freq_vowel_AcAttn}
    \vspace{-0.15cm}
\end{figure*}
\subsection{Mean Time-Frequency Representations}
We analyze the learned time-frequency representation of each phoneme from the first layer (using the representation $\boldsymbol{z}$ with $7$ previous and $7$ succeeding frames that cover $165$ ms duration). For example, the time-frequency representation of all /aa/ vowel exemplars (denoted as $\boldsymbol{z}$ in Figure~\ref{fig:block_diag}) are extracted from TIMIT utterances and are averaged to obtain one average time-frequency representation, as shown in Figure \ref{fig:avg_phoneme_spec} (top row second column).  \textcolor{black}{For the computation of the average spectrogram for each phoneme, a contextual window of $+/- 7$ frames are chosen around the center frame (from every exemplar of that phoneme occurrence in the files) irrespective of the phone duration. Thus, all center frames belonging to a phoneme, except the two boundary frames on either side, are used in computing the average time-frequency representation of Figure~\ref{fig:avg_phoneme_spec}.}
%The yellow areas indicate regions of higher average energy and black regions indicate weaker average energy. 

The averaged time-frequency representation learned by the model reveals that mid/back vowels \{/ao/, /aa/, /ae/\} have relatively more concentrated activity at low to medium frequencies ($0.5-2$ KHz), whereas front vowels \{/ey/, /iy/\}  have two distinct peaks spaced over a larger frequency range around $0.3$ and $4$ kHz respectively. This is consistent with the known distribution of the three formants F1, F2, and F3 in these vowels \cite{mesgarani2008phoneme}. %As the vowel articulation becomes more forward, the single peak broadens and splits /ae/ to /ey/. Continuing this trend, Front/Closed vowels \{/iy/, /u/\} exhibit relatively narrow formant peaks with F1 at low frequencies.
On the other hand, while the plosive /t/,  being a stop-consonant,  has a time varying profile, the fricatives \{/sh/, /zh/\} have dominant energy at high frequencies and the nasal sounds  \{/m/, /em/\} have energy at low to mid frequencies. Thus, the early representations learned by the model attempt to capture distinct phonetic properties. {\textcolor{black}{The purpose of this analysis is to relate the relevance weight plots that are analyzed in the following sections to the phonetic properties of the time-frequency representation.}}

\begin{figure}[t]
    \centering
    \includegraphics[trim={0 0.3cm 0 0.9cm}, clip, scale=0.38]{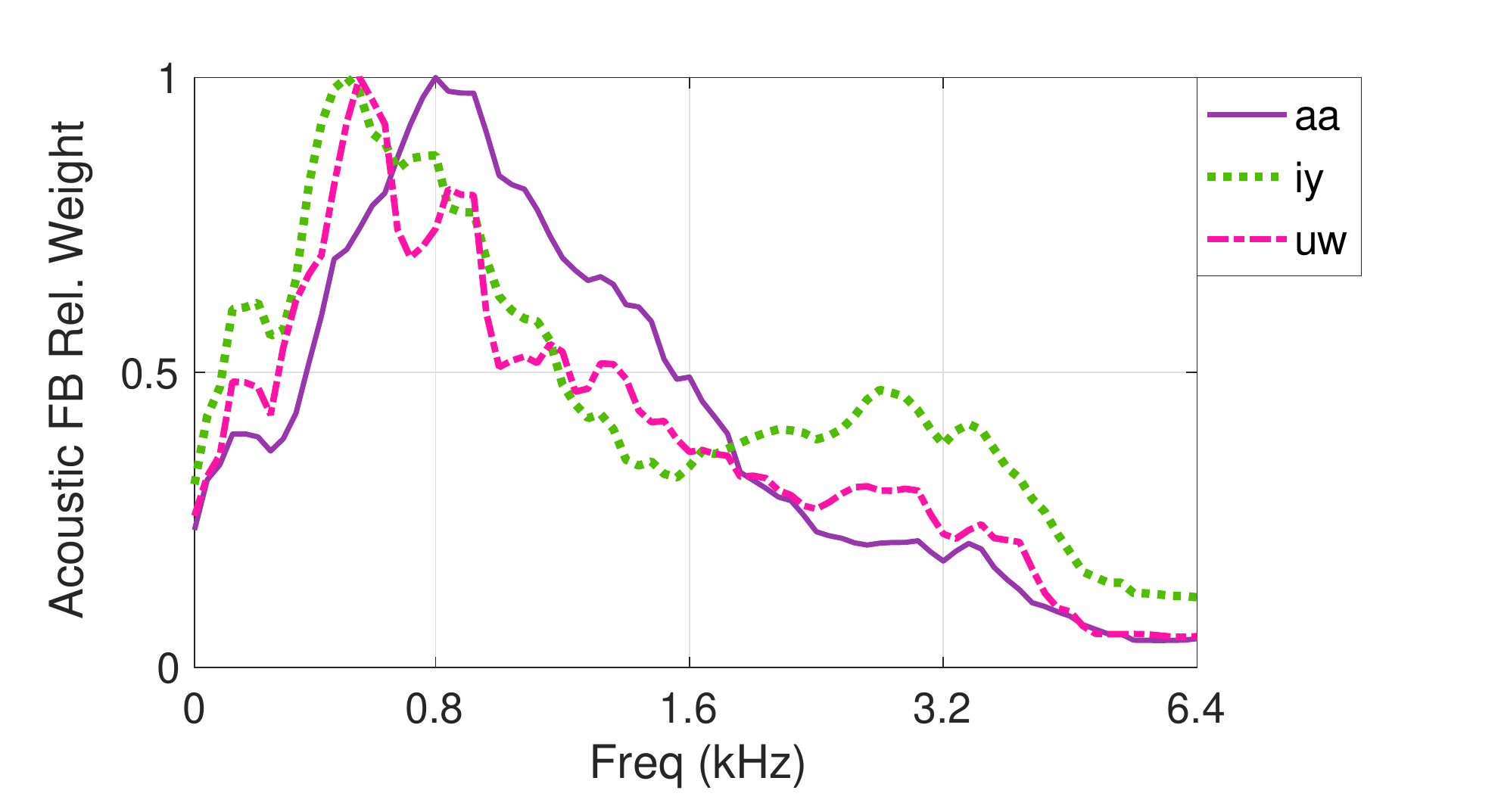}
    \vspace{-0.2cm}
    \caption{{Vowel Analysis - Acoustic filterbank (FB) relevance weights for $3$ vowels on clean TIMIT data (black dotted curve for vowels in Figure} \ref{fig:best_freq_vowel_AcAttn}). This figure highlights the contrast among vowels for clean condition.}
    \label{fig:acAttn_3vowels_cleanTIMIT}
    \vspace{-0.25cm}
\end{figure}
\vspace{-0.2cm}
\subsection{Acoustic FB Relevance Weights}
The acoustic filterbank relevance weights $\boldsymbol{w_a}$ are analyzed  to understand the weighting incorporated through the relevance sub-network. The relevance weights are averaged across the utterances for each phoneme 
%of each noisy type 
and SNR level separately. Figure \ref{fig:best_freq_vowel_AcAttn} shows the relevance weights \textcolor{black}{for clean data and noisy data averaged over all the $4$ noise types}. 
%Without relevance weighting, it can be assumed that each sub-band weight is be equal to $1/80 = 0.0125$.

As can be observed for \textcolor{black}{clean and} SNR of $20$ dB (low noise), for the front vowels \{/ey/, /iy/, /uw/\}, the filter indices that have the higher relevance weights lie in $500-600$ Hz range. For the mid/back vowels \{/ao/, /aa/, /ae/\}, more relevance is seen in the filter indices having center frequency above $800$ Hz. The filter indices with the highest relevance weights \textcolor{black}{(peak marked with black star in Figure \ref{fig:best_freq_vowel_AcAttn})} tend to shift towards lower frequency as we move from the back vowel /ao/ to the front vowel /uw/. This is also very similar to the filter representations seen in the auditory system of ferrets and other mammals \cite{mesgarani2008phoneme}. Also, in front/closed vowels \{/iy/, /uw/\}, the relevance weights show a second peak at the higher frequency index due to the presence of a higher formant frequency F3 (in the center frequency range of $3.5-4$ kHz). For the high noise condition (SNR $0$ dB), the relevance weights still preserve some of the phoneme specific selectivity although the weights tend to become more uniform across the frequency range. \textcolor{black}{A more closer look at the relevance weights is given in Figure~\ref{fig:acAttn_3vowels_cleanTIMIT}, where $3$ vowels are analyzed in clean conditions. As seen here, the peak activity moves from $800$Hz to $500$Hz as we move from back vowels to front vowels.}

The acoustic filterbank (FB) relevance weights for consonants is shown in the second row of Figure \ref{fig:best_freq_vowel_AcAttn}. The plosive /t/ shows similar relevance weighting as observed in front vowels. The fricatives \{/sh/, /zh/\} show lesser sub-band selectivity compared to other phonemes (as these phonemes have significant high frequency activity as seen in Figure~\ref{fig:avg_phoneme_spec}). The nasal sounds \{/em/, /eng/\} also elicit a large range of filter indices that have high  relevance weights.
Similar to the vowels, the decrease in SNR (higher noise case) reduces the selectivity of the representations as the  relevance weights become more uniform. This is again similar to the  human auditory processing where noise segregation happens to a very small degree in the peripheral time-frequency representation as compared to higher auditory cortical areas \cite{da2013tuning}.  
%Also, in all the phonemes with SNR $0$, the relevance sub-band weights are higher for $10$ to $65$ numbered sub-bands, i.e. $0.4$ to $6$ kHz as compared to other sub-band regions.
% Among the 4 noise types, babble, exhall and restaurant have higher weights compared to subway noise in all vowels.
% Since the general trend of the 4 noise types (babble, exhall, restaurant and subway) appear to be similar, we will analyze the average of the 4 noise types in rest of the analysis.
\vspace{-0.15cm}
\subsection{Modulation Filtering With Relevance Weighting}
\begin{figure*}[t]
    \centering
    % \vspace{-0.1cm}
    \hspace{0.4cm}
    \includegraphics[trim={2cm 0cm 0 0}, clip, scale=0.41]{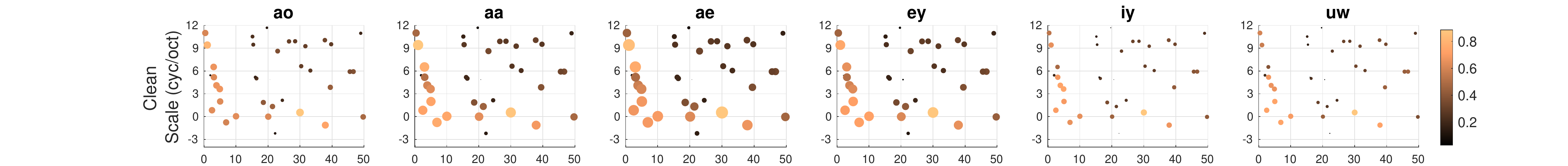}\\\vspace{0.15cm}
    \includegraphics[trim={0 0 0 0}, clip, scale=0.4]{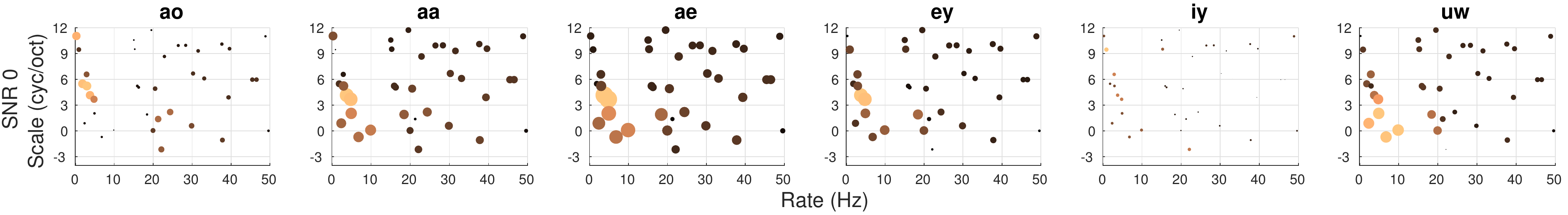}\\\vspace{0.15cm}
    % \hspace{0.05cm}
    \includegraphics[trim={2cm 0cm 0 0}, clip, scale=0.395]{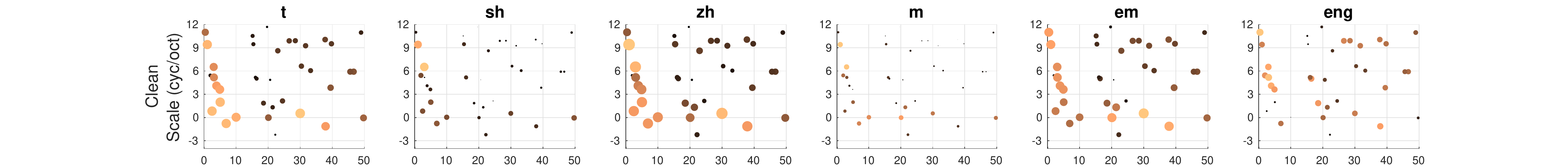}\\\vspace{0.15cm}
    % \hspace{0.05cm}
    \includegraphics[trim={0 0 0 0}, clip, scale=0.4]{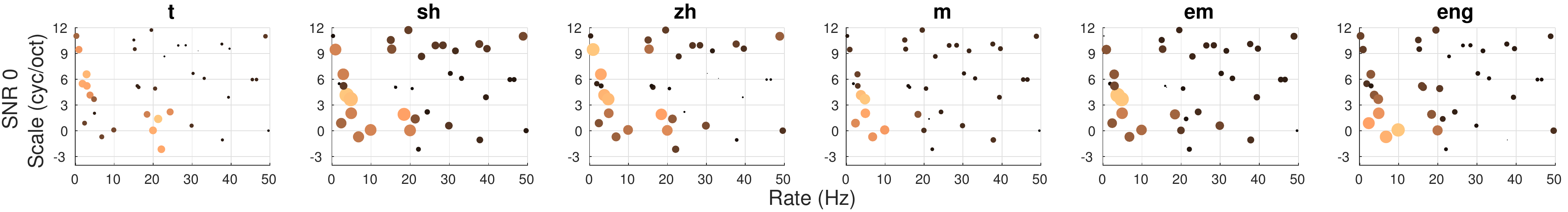}
    \vspace{-0.2cm}
    \caption{The modulation relevance weights (after removing the mean weights)  plotted for each phoneme:  \textcolor{black}{vowel phonemes in top two rows for clean and  SNR $0$ dB condition and consonants in the last two rows for clean and SNR $0$ dB condition respectively}. The size of the bubble is proportional to the magnitude of the relevance weight.}
    \label{fig:rate_scale_modAttn}
    \vspace{-0.45cm}
\end{figure*}

Figure \ref{fig:rate_scale_modAttn} shows the bubble plot for modulation filter relevance weights for different phonemes (averaged over all respective exemplars of each phoneme). This plot is obtained by placing the relevance weight at the location of the center frequency (rate-scale) of the corresponding modulation filter. The size of the bubble is proportional to the magnitude of the corresponding relevance weight. The vowels and consonants are arranged in the same order as in acoustic FB relevance weight analysis. 
In order to highlight the contrast, we plot relevance weights for a given phoneme after subtracting the mean relevance weights over all the $50$ TIMIT phonemes. 
%Hence, a low relevance weight in the plot indicates lower difference from the mean relevance weight.

The top row shows the weights for vowels \textcolor{black}{under clean condition. It can be observed that almost all the vowels have higher contrast in relevance weight values for the rate frequency in $0-12$ Hz rate and $-1$ to $9$ cyc./oct scale. Additionally, most of the back and mid vowels elicit higher contrast in weights for band-pass rate and low-pass scale frequencies.} 
% The vowel /iy/ has almost all relevance weights close to mean relevance weights, and hence very small bubble sizes. 
In the low SNR condition, for vowels (second row of Figure \ref{fig:rate_scale_modAttn} with SNR$=0$ dB), the relevance characteristics are much more intact compared to the acoustic FB layer relevance weights, with more contrast  in lower rate regions.
% Thus, the second layer of relevance weights  are more resilient to the presence noise for most of the vowels.
In case of consonants (third and fourth row of Figure \ref{fig:rate_scale_modAttn}), for the \textcolor{black}{clean case, the plosive /t/ show similar trend as mid vowels while the fricative /sh/ show a low-pass rate + high-pass scale contrast profile. Additionally, most of the consonants have high contrast towards low-pass rate + high-pass scale and band-pass rate + low-pass scale region. On the other hand, the noisy consonants (SNR 0 dB) show low-pass rate profile scattered over different scale values.}
%At high levels of noise (SNR of $0$ dB), the modulation relevance weights for the consonants show more noise resilience compared to the acoustic FB counterparts.

%the low and medium scale frequency in all the consonants (plosive, fricatives, nasals) with band-pass rate characteristic in plosive, fricatives and low-pass+band-pass rate in nasal phonemes. in high noise condition with SNR$=0$, the rate characteristics tend to move towards low rate in all the consonants. It can be attributed to the relevance weights trying to capture underlying phonetic content while also enabling the suppression of noise.

\vspace{-0.2cm}
\section{Experiments and results}{\label{sec:experiments}}
The speech recognition system is trained using PyTorch toolkit \cite{paszke2017pytorch} while the Kaldi toolkit~\cite{povey2011kaldi} is used for decoding and language modeling. 
% The ASR is built on three datasets, Aurora-4, CHiME-3 and VOiCES respectively. 
The models are discriminatively trained using the training data with cross entropy loss and Adam optimizer \cite{kingma2014adam}. A hidden Markov model - Gaussian mixture model (HMM-GMM) system is used to generate the senone alignments for training the CNN-DNN based model.
% The ASR results are reported with a language model re-scoring of the lattices, where the lattices generated with tri-gram language model are rescored using recurrent neural network language model (RNN-LM) \cite{mikolov2011rnnlm} for the final ASR decoding.
The ASR results are reported with a tri-gram language model and the best language model weight is obtained from the development set. 
\textcolor{black}{The notation [A,M] refers to the model of learning acoustic and modulation filterbanks without relevance, [A-R,M] refers to the model that involves acoustic FB with relevance along with modulation FB without relevance, [A, M-R] refers to learning acoustic FB (with no relevance weighting), followed by modulation filter learning with relevance weighting, and [A-R,M-R] refers to the model with learning acoustic and modulation FB and having relevance weighting in both layers. The modulation filters are learnt in all baseline features as well as the first 2-D CNN layer.}

For each dataset, we compare the ASR performance of the proposed approach with traditional mel filterbank energy (MFB) features, power normalized filterbank energy (PFB) features \cite{kim2012pncc}, RASTA features that perform modulation filtering (RAS) \cite{hermanskyb}, and mean Hilbert envelope (MHE) features \cite{mhec2015}. {All the baseline features are processed with cepstral mean and variance normalization (CMVN) on a $1$ sec. running window}.  {\textcolor{black}{The baseline MFB features are directly fed to the 2-D modulation filtering layer (without the acoustic FB layer). The other baseline features like PFB, RAS and MHE also generate spectrogram like time-frequency representations which are used similar to the MFB features. For all these features, no relevance weighting is performed.  The modulation filtering layer (M) is part of the baseline system and is present with all the other features like PFB, RAS, MHE (without explicit mention in all cases). An additional experiment, where relevance weighting is applied on the baseline MFB features (denoted as MFB-R) is also performed.}} 
% The neural network architecture shown in Figure \ref{fig:block_diag} (except for the acoustic filterbank learning layer, the acoustic FB relevance sub-network and modulation filter relevance  sub-network) is used for all the baseline features. 
For ASR experiments on the proposed model, we use a non-parametric approach in modulation filter learning. \textcolor{black}{The non-parametric approach to modulation filtering involves learning 2-D kernels of the CNN layer that operates on the output of the acoustic FB layer.} Empirically, the non-parametric modulation filters had a slight improvement over parametric modulation filters in ASR performance. The batch size of $32$ is chosen for all the model training using a learning rate of $= 0.0001$. The model training is performed for $10$ epochs after which the learning is found to saturate on the validation data.

  \begin{center}   
     \begin{table}[t]
     %\vspace{-0.2cm}
            \centering
            \begin{center}
            \caption{Word error rate (\%) in Aurora-4 database for multi-condition training with various feature extraction schemes.}
            \label{tab:multiData_aurora4}
            %\vspace{-0.3cm}
            % \resizebox{8.8cm}{3.5cm}{
            \begin{tabular}{l|c|c|c|c|c}
            \hline
            Cond & MFB,M & PFB,M  & RAS,M & {MHE,M} & A-R,M-R\\ \hline
            \multicolumn{6}{c}{\textcolor{black}{Aur-A} : Clean with same Mic} \\ \hline
            Clean & 4.2 & 4.0 & 5.3 & 3.8 & \textbf{3.6}\\ \hline
            \multicolumn{6}{c}{\textcolor{black}{Aur-B} : Noisy with same Mic} \\ \hline
            Airport & 6.8 & 7.1 & 8.1 & 7.3 & \textbf{5.9} \\
            Babble & 6.6 & 7.4 & 8.0 & 7.4 & \textbf{6.1} \\
            Car & {4.0} & 4.5 & 4.8 & 4.3 &  \textbf{3.9} \\
            Rest. & 9.4 & 9.6 & 11.0 & 9.1 & \textbf{6.8} \\
            Street & 8.1 & 8.1 & 9.4 & 7.6 & \textbf{6.9} \\
            Train & 8.4 & 8.6 & 9.6 & 8.6 & \textbf{7.2} \\\hdashline
            Avg. & 7.2 & 7.5 & 8.5 & 7.4 & \textbf{6.1}\\
             \hline
            \multicolumn{6}{c}{\textcolor{black}{Aur-C} : Clean with diff. Mic}\\ \hline
		    Clean & 7.2 & 7.3 & 9.0 & 7.3 & \textbf{6.0} \\ \hline
            \multicolumn{6}{c}{\textcolor{black}{Aur-D} : Noisy with diff. Mic}  \\ \hline
            Airport & 16.3 & 18.0 & 17.5 & 17.6 &  \textbf{14.1} \\
            Babble & 16.7 & 18.9 & 19.0 & 18.6 &   \textbf{15.4}\\
            Car & {8.6} & 11.2 & 10.5 & 9.6 & \textbf{7.7}\\
            Rest. & 18.8 & 21.0 & 21.3 & 20.1 & \textbf{18.6} \\
            Street & 17.3 & 19.5 & 18.5 & 18.8 & \textbf{16.8}\\
            Train\ & 17.6 & 18.8 & 19.4 & 18.7 & \textbf{16.2} \\ \hdashline
            Avg. & 15.9 & 17.9 & 17.7 & 17.3 & \textbf{14.8} \\
             \hline
            \multicolumn{6}{c}{Avg. of all conditions}  \\ \hline
            Avg. & 10.7 & 11.7 & 12.2 & {11.4} & \textbf{9.6}\\ \hline
            \end{tabular}
            % }
          \end{center}
          \vspace{-0.25cm}
      \end{table}
  \end{center}
\vspace{-0.45cm}
\subsection{Aurora-4 ASR}
%The WSJ Aurora-4 corpus is used for conducting ASR experiments. 
Aurora-4 database consists of continuous read speech recordings of $5000$ words corpus, recorded under clean and noisy conditions (street, train, car, babble, restaurant, and airport) at $10-20$ dB SNR. The training data has $7138$ multi condition recordings ($84$ speakers) with total $15$ hours of training data. The validation data has $1206$ recordings for multi condition setup. The test data has $330$ recordings ($8$ speakers) for each of the $14$ clean and noise conditions. The test data are classified into group \textcolor{black}{Aur-A - clean data, Aur-B - noisy data, Aur-C - clean data with channel distortion, and Aur-D} - noisy data with channel distortion.

The ASR  performance on the Aurora-4 dataset is shown in Table \ref{tab:multiData_aurora4} for each of the $14$ test conditions. As seen in the results, most of the noise robust front-ends do not improve over the baseline mel filterbank (MFB) performance. The proposed representation learning (two-stage relevance weighting)  provides  considerable improvements in ASR performance over the baseline system with average relative improvements of $11$\% over the baseline MFB features. Furthermore, the improvements in ASR performance are consistently seen across all the noisy test conditions. 
%In particular, the relative improvements in same microphone conditions (A and B) are about $15$\% relative compared to the baseline system. 
\subsubsection{Statistical Significance}
In order to compare how one system performs relative to the other in statistical sense, we use the bootstrap estimate for confidence interval \cite{bisani2004bootstrap}. This method computes a bootstrapping of word error rate (WER) values to extract the $95$\% confidence interval (CI), and also gives a probability of improvement (POI) for the system-in-test (system with the proposed representation learning) over the reference system (baseline system with MFB features). Table \ref{tab:multiData_aurora4_bootstrap_POI} shows the analysis for various test conditions in the Aurora-4 multi-condition training.
% The bootstrap estimate of CI is similar for MFB and our proposed method. 
The POI of {\textcolor{black}{the}} proposed system (A-R,M-R) system over the MFB is quite high for all the test conditions, with average POI being  $94$\%.
\begin{center}
\begin{table}[t]
    \centering
    \caption{Statistical significance of performance improvements for the proposed method over the baseline MFB system using confidence interval and the probability of improvement (POI) on Aurora-4 dataset \cite{bisani2004bootstrap}.  
    %are reported below for the various conditions in the Aurora-4 dataset. 
    }
\label{tab:multiData_aurora4_bootstrap_POI}
\resizebox{7.1cm}{1.25cm}{
    \begin{tabular}{l|c|c|c}
    \hline
    Test Cond. & \multicolumn{2}{c|}{Confidence Interval} & POI (\%)\\\hline
    & MFB,M & A-R,M-R & \\\hline
    \textcolor{black}{Aur-A} & [4.1, 5.3 ] & [ 3.8, 5.0 ] & 95.1 \\
    \textcolor{black}{Aur-B} & [ 7.3, 9.7 ] & [ 7.2, 9.4 ] & 86.8 \\
    \textcolor{black}{Aur-C} & [ 7.7, 10.7 ] & [ 6.5, 9.1 ] & 99.0 \\
    \textcolor{black}{Aur-D} & [ 17.4, 23.0 ] & [16.4, 21.8 ] & 95.3 \\\hline
    Avg &-- & --& 94.0 
    \\\hline
    \end{tabular}
}
%\vspace{-0.3 cm}
\end{table}
\end{center}
\begin{table}[t]
\begin{center}
\caption{Word error rate (\%) in CHiME-3 Challenge database for multi-condition training (real+simulated) with test data from simulated and real noisy environments.}
% \vspace{0.05in}
\label{tab:Chime3Results}
\vspace{-0.3cm}
    \resizebox{\columnwidth}{!}{
	\begin{tabular}{l|c|c|c|c|c|c}
	\hline
		Cond. & MFB,M & PFB,M & {RAS,M} & {MHE,M} & A-R,M & A-R,M-R \\ \hline
		\multicolumn{7}{c}{Dev}  \\ \hline  
	%	\multicolumn{8}{|c|}{Multi condition reverb training} \\ \hline
		Sim & 12.9 & 13.3 & 14.7 & 13.0 & {12.5} & \textbf{12.0}\\ 
		Real & 9.9 & 10.7 & 11.4 & 10.2 & {9.9}  & \textbf{9.6}\\ \hdashline
		Avg. & 11.4 & 12.0 & 13.0 & 11.6  & {11.2}  & \textbf{10.8}\\ \hline
		
		\multicolumn{7}{c}{Eval} \\ \hline 

		Sim & 19.8 & 19.4 & 22.7 &  19.7 & {19.2} &  \textbf{18.5}\\ 
		Real & 18.3 & 19.2 & 20.5 & 18.5  & {17.3}  & \textbf{16.6}\\\hdashline
		Avg. & 19.1 & 19.3 & 21.6 &  19.1 & {18.2}  & \textbf{17.5}\\ \hline 

% 		Best overall (dev)& 13.0 &  12.8 & {13.2} & {13.2} & 11.4 & \textbf{11.3} \\ \hline
	\end{tabular}
	}
\end{center}
\vspace{-0.3cm}
\end{table}
\vspace{-0.8cm}
\subsection{{CHiME-3 ASR}}
The CHiME-3 corpus for ASR contains multi-microphone tablet device recordings from everyday environments, released as a part of 3rd CHiME challenge \cite{barker2015chime3}. Four varied environments are present - cafe (CAF), street junction (STR), public transport (BUS) and pedestrian area (PED). For each environment, two types of noisy speech data are present - real and simulated. The real data consists of $6$-channel recordings of sentences from the WSJ$0$ corpus spoken in the environments listed above. The simulated data was constructed by artificially mixing clean utterances with environment noises. The training data has $1600$ (real) noisy recordings and $7138$ simulated noisy utterances, constituting a total of $18$ hours.
% The real data is supplemented by $7138$ simulated utterances constructed by taking the full WSJ0 5k training set and mixing it with the separately recorded CHiME-3 noise backgrounds.
We use the beamformed audio in our ASR training and testing. The development (dev) and evaluation (eval) data consists of $410$ and $330$ utterances respectively. For each set, the sentences are read by four different talkers in the four CHiME-3 environments. This results in $1640$ ($410 \times 4$) and $1320$ ($330 \times 4$) real development and evaluation utterances. 
%Identically-sized, simulated dev and eval sets are made by mixing recordings captured in the recording booth with the environmental noise recordings.

The results for the CHiME-3 dataset are reported in Table \ref{tab:Chime3Results}. The approach of acoustic FB learning with relevance weighting alone (A-R) improves over the baseline system (MFB) as well as the other noise robust front-ends considered here. The proposed approach of 2-stage relevance weighting over learned acoustic representations and modulation representations (A-R,M-R) provides significant improvements over baseline features. On the average, the proposed approach provides relative improvements of $10$\% over MFB features in the eval set. 
The detailed results on different noises in CHiME-3 are reported in Table \ref{tab:Chime3Results_detailed}. For most of the noise conditions in CHiME-3 in simulated and real environments, the proposed approach provides consistent improvements over the baseline features.

\begin{table}[t]
\begin{center}
\caption{WER (\%) for each noise condition in CHiME-3 dataset with the baseline features and the proposed feature extraction.}
\label{tab:Chime3Results_detailed}
\vspace{-0.3cm}
    \resizebox{\columnwidth}{!}{
	\begin{tabular}{|c|c|c|c|c|c|c|c|c|}
	\hline
	 & \multicolumn{4}{c|}{\textbf{Dev Data}} & \multicolumn{4}{c|}{\textbf{Eval Data}} \\\hline
	\multirow{2}{*}{Cond.} & \multicolumn{2}{c|}{MFB,M} & \multicolumn{2}{c|}{A-R,M-R} & \multicolumn{2}{c|}{MFB,M} & \multicolumn{2}{c|}{A-R,M-R} \\\cline{2-9}
	& Sim & Real & Sim & Real & Sim & Real & Sim & Real \\\hline
	BUS & \textbf{10.9} & 11.6 & \textbf{10.9} &  \textbf{{11.3}} & 13.7 & 22.5  & \textbf{13.0} &  \textbf{21.4}\\
	CAF & 16.8 & 9.8 & {\textcolor{black}{\textbf{14.7}}} &  \textbf{9.5} & 22.3 & 18.8 & \textbf{19.6} &  \textbf{16.1} \\
	PED & 10.4 & 8.0 & \textbf{9.6} &  \textbf{7.5} & 20.8 & 17.7 & \textbf{18.7} &  \textbf{15.4}\\
	STR & 13.8 & 10.3 & \textbf{13.0} & \textbf{10.0} & \textbf{22.5} & 14.4 & {22.7} &  \textbf{13.3} \\ \hline
	\end{tabular}
	}
\end{center}
\vspace{-0.15cm}
\end{table}
\begin{center}
    \begin{figure}[t]
        \centering
        \includegraphics[trim={0 0 0 0cm}, clip, scale=0.32]{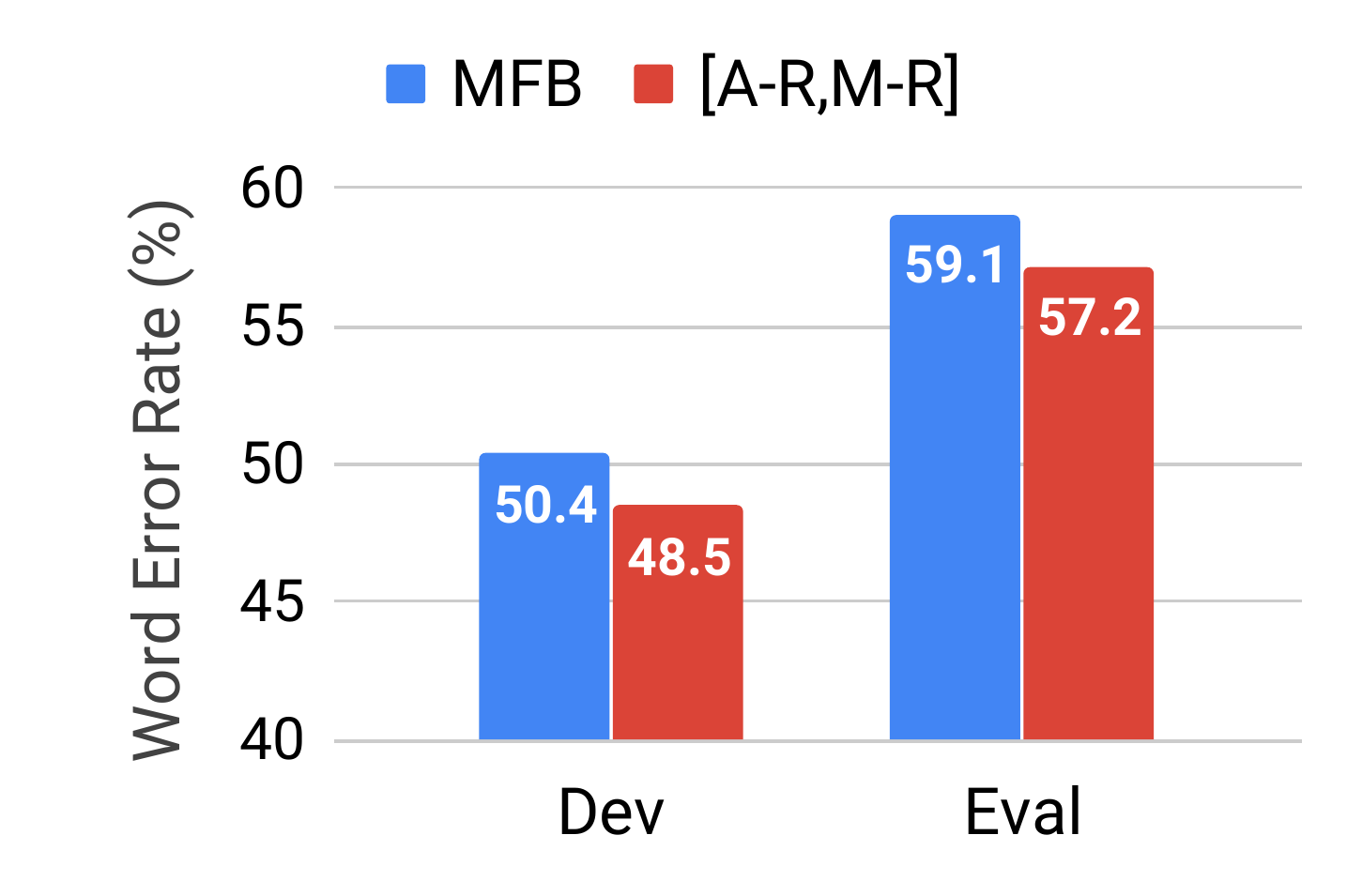}
        \vspace{-0.2cm}
        \caption{ASR performance in WER (\%) for VOiCES database.}
        \label{fig:voices_asr}
        %\vspace{-0.1cm}
    \end{figure}
\end{center}
\vspace{-0.9cm}
\subsection{VOiCES ASR}
The Voices Obscured in Complex Environmental Settings (VOiCES) corpus is a creative commons speech dataset \cite{richey2018voices}, being used as part of VOiCES Challenge \cite{nandwana2019voices}.
% The ASR training data under 'fixed' training condition track contains a subset of Librispeech clean training data in this challenge.
The training data set of $80$ hours has $22,741$ utterances sampled at $16$kHz from $202$ speakers, with each utterance having $~12-15$s segments of read speech from the Librispeech corpus. 
%This subset was designed in such a way as to have no overlap in speakers with the VOiCES corpus (development or evaluation). 
We performed a 1-fold reverberation and noise augmentation of the data using Kaldi \cite{povey2011kaldi}.
The ASR development set consists of $20$ hours of  distant recordings from the $200$  VOiCES dev speakers. It contains recordings from $6$ microphones. The evaluation set  consists of $20$ hours of distant recordings from the $100$ VOiCES eval speakers and contains recordings from $10$ microphones from an unseen room. The ASR performance of VOiCES dataset with baseline MFB features and our proposed approach of 2-step relevance weighting is reported in Figure \ref{fig:voices_asr}. 
%Our experiment over this large dataset is with the motive to assess the scalability of the proposed method (the baseline ASR performance can definitely be improved with better baseline models). 
These results suggest that the proposed model is also scalable to relatively larger ASR tasks with large vocabulary where consistent improvements are obtained with the proposed approach in addition to the interpretability of the representations learned. %\vspace{-0.9cm}
%From the results, it can be observed that the proposed representation learning approach performs better than the baseline MFB features for both Dev and Eval test data.

\begin{center}
    \begin{table}[t]
        \centering
        \caption{Effect of relevance weighting on different stages of the proposed model on ASR with Aurora-4 dataset.
        }
        \label{tab:effect_relevanceWt_ASR}
        \vspace{-0.1cm}
        % \resizebox{7.5cm}{0.8cm}{
        \begin{tabular}{l|c|c|c|c|c}
        \hline
         \multirow{2}{*}{\textbf{Features}} & \multicolumn{5}{c}{\textbf{ASR (WER in \%)}} \\ \cline{2-6}
          & \textcolor{black}{Aur-A} & \textcolor{black}{Aur-B} & \textcolor{black}{Aur-C} & \textcolor{black}{Aur-D} & Avg.  \\\hline
         MFB,M & 4.2 & 7.2 & 7.2 & 15.9 & 10.7\\
         A,M (Acoustic FB) & 4.1 & 6.8 & 7.3 & 16.2 & 10.7 \\
         MFB-R,M & 4.0 & 7.3 & 7.1 & 16.1 & 10.8 \\
         A-R,M & 3.6 & 6.4 & 8.1 & 15.1 & 10.0\\
         {A,M-R} & \textbf{3.4} & \textbf{6.1} & 7.9 & 16.9 & 10.4\\
         \textcolor{black}{MFB-R, G-R} & 3.7 & 6.3 & 6.4 & 15.1 & 10.3 \\
         A-R,M-R & 3.6 & \textbf{6.1} & \textbf{6.0} & \textbf{14.8} & \textbf{9.6}\\
         \hline
        \end{tabular}
        % }
        % \vspace{-0.1cm}
    \end{table}
\end{center}
\begin{center}
    \begin{table}[t!]
        \centering
        \caption{WER (\%) for cross-domain representation learning and ASR training experiments \textcolor{black}{with all $3$ datasets. The performance with MFB (fixed mel filterbank) features is also reported in parenthesis for reference in first row}.
        }
        \label{tab:cross_domain_ASR}
        \vspace{-0.1cm}
        % \resizebox{7.5cm}{0.8cm}{
        \begin{tabular}{l|c|c|c}
        \hline
         \multirow{2}{*}{\textbf{[A-R,M-R] Learned on }} & \multicolumn{3}{c}{\textbf{ASR Trained and Tested on}} \\ \cline{2-4}
          & Aurora-4 & CHiME-3 & VOiCES  \\\hline
         Aurora-4 & \textbf{9.6} \textcolor{black}{(10.7)} & 14.3 \textcolor{black}{(15.3)} & 57.6 \textcolor{black}{(59.1)}\\
         CHiME-3 & 9.7 & \textbf{14.2} & 59.0 \\
         VOiCES & 9.9 & 14.4 & \textbf{57.2} \\\hline
        \end{tabular}
        % }
        \vspace{-0.2cm}
    \end{table}
\end{center}
\vspace{-1.4cm}
\section{{Discussion}}\label{sec:discussion}
\subsection{Effect of relevance weighting on different stages}
% Here we analyze the effect of relevance weighting on different stages of the proposed model. 
We compare the ASR performance of the proposed approach of 2-stage relevance weighting (A-R,M-R) with the learned acoustic FB representation (A) without any relevance weighting, relevance weighting on \textcolor{black}{fixed} mel filterbank features (MFB-R), acoustic FB relevance weighting without modulation FB relevance weighting (A-R), and modulation relevance weighting alone on learned time-frequency representation (A,M-R). \textcolor{black}{We also report performance with handcrafted filterbanks at both the stages, denoted as [MFB-R, G-R]. The hand-crafted modulation filters (G) were chosen as 2-D cosine-modulated Gaussian filters with pre-determined rate-scale center frequencies. Based on prior studies, we use more filters in the band-pass rate and low pass scale regions.}

The results are reported in Table \ref{tab:effect_relevanceWt_ASR} averaged for $4$ test conditions of Aurora-4 dataset.
\textcolor{black}{Since the modulation filtering layer (M) is part of the baseline system and is present with all the features, the notation `M' alone is omitted in discussion.} 
As can be observed, the acoustic FB features (A) performs similar to the MFB baseline features on average. The MFB-R features, which denote the application of the relevance weighting over mel filterbank features, does not provide improvements over baseline MFB features. The relevance weighted features A-R improves over the acoustic FB (A) features with average relative improvements of $6$\%. The features (A,M-R) having modulation relevance weighting alone improves over baseline in condition Aur-A and Aur-B, while there is degradation in Aur-C and Aur-D. {\textcolor{black}{The handcrafted features (MFB-R,G-R) also improves over the baseline}}. The proposed 2-stage relevance weighting (A-R,M-R) provides improvements in all test conditions over the baseline.  

\subsection{Representation transfer across tasks}
% In a subsequent analysis, we perform a cross-domain ASR experiment, i.e., we use the acoustic filterbank learned from one of the datasets (either Aurora-4 or CHiME-3 challenge)  to train/test ASR on the other dataset.
In a subsequent analysis, we perform a cross-domain ASR experiment, i.e., we use the proposed representation learning (A-R,M-R)
% (acoustic FB + acoustic FB relevance weight + modulation FB + modulation FB relevance weight) 
learned from one of the datasets (either Aurora-4, CHiME-3 or VOiCES challenge)  to train/test ASR on the other dataset. All other layers in Figure~\ref{fig:block_diag} are learned using the in-domain ASR.
The results of these cross-domain representation learning and ASR training experiments are reported in Table~\ref{tab:cross_domain_ASR}.
% The rows in the table show the database used to learn the representations and the columns show the dataset used to train and test the ASR.
The performance reported in this table are the average WER on each of the datasets. The results shown in Table~\ref{tab:cross_domain_ASR} illustrate that the representation learning process is relatively robust to the domain of the training data, which suggest that the proposed representation learning approach can be generalized for other ``matched'' tasks (especially between Aurora-4 and VOiCES tasks).
%without the need to re-learn the filterbank parameters. 

\begin{table}[t]
    \centering
    \caption{Comparison of MFB with different \textcolor{black}{learnable front-end} methods - without and with relevance weighting \textcolor{black}{on Aurora-4 dataset}. }
    \label{tab:sincNet}
    \vspace{-0.1cm}
    \begin{tabular}{c|c}
    \hline
        Features without relevance weighting & ASR \\ \hline
        MFB,M (mel filterbank) & 10.7 \\
        A,M (our cosine modulated Gaussian filterbank) & 10.7 \\
        Sinc,M (sinc filterbank from \cite{ravanelli2018interpretable}) & 10.8\\\hline \hline
    \end{tabular}
% \end{table}
% \begin{table}[t]
%     \centering
%     \caption{Comparison of MFB with different filterbank learning methods in proposed relevance weighting based framework.}
%     \label{tab:sincNet_attn}
    \begin{tabular}{c|c}
    % \hline
        Features with relevance weighting & ASR \\ \hline
        MFB-R,M-R & 10.6 \\
        A-R,M-R & \textbf{9.6} \\
        Sinc-R,M-R &  10.0 \\
        \textcolor{black}{ANP-R,M-R} & 11.6 \\
        \textcolor{black}{A-SelfAttn,M} & 10.8 \\\hline 
    \end{tabular}
    \vspace{-0.15cm}
\end{table}
\vspace{-0.15cm}
\subsection{Comparison with other \textcolor{black}{learnable front-end methods}}
% One of the popular existing method to learn interpretable acoustic filterbank from the raw data in a supervised learning paradigm is SincNet \cite{ravanelli2018interpretable}. This method uses a convolution layer as first layer to learn sincfilters as parametric acoustic filters from raw waveform, with only low and high cutoff frequencies of band-pass filters to be directly learned from data.

To compare the proposed approach in this work with the SincNet method \cite{ravanelli2018interpretable}, we replace our cosine modulated Gaussian filterbank with the sinc filterbank as kernels in first convolutional layer. The baseline ASR system with sinc filterbank
%\footnote{https://github.com/mravanelli/SincNet/} 
is trained jointly without any relevance weighting and the rest of the architecture is kept the same as shown in Fig. \ref{fig:block_diag}. The ASR system with sincFB and 2-stage relevance weighting is also trained. {\textcolor{black}{In addition, we also compare with acoustic FB learning in a non-parametric (ANP) manner (learning the 1-D CNN kernels directly from raw input without using a cosine modulated Gaussian function) \cite{doss2013, tuske2014acoustic, sainath2015cldnn}}}.

The ASR performance is reported in Table \ref{tab:sincNet} for mel filterbank features (MFB), our proposed cosine modulated Gaussian filterbank without relevance weighting (R) and the sinc filterbank from \cite{ravanelli2018interpretable} (Sinc). For the experiments without relevance weighting, it can be observed that the parametric sinc filterbank performs similar to our acoustic filterbank, and both perform similar to mel filterbank. The relevance weighting over sinc FB (Sinc-R,M-R) improves over the baseline with average relative improvement of around $6$\% over MFB, while the proposed [A-R,M-R] representation improves over the sinc FB approach. \textcolor{black}{The proposed parametric approach to acoustic FB learning also improves over the non-parametric (ANP) approach.}

\textcolor{black}{In addition, to compare the relevance weighting approach with the self-attention approach, we train an ASR model with the acoustic FB relevance sub-network   replaced with self-attention module (with self-attention weighting on sub-bands) for the acoustic filterbank layer. From the ASR performance for this system (A-SelfAttn,M), it can be observed that the relevance weighting proposed in this work is superior to channel weighting based on self-attention.}

\subsection{Unsupervised vs. Supervised representation learning}
In our proposed approach, the parametric kernels of the acoustic FB layer
% and kernels of the modulation filtering layer 
can be initialized in different ways; (i) initialization using unsupervised training of CVAE \cite{agrawal2019unsupervised}, (ii) unsupervised initialization + supervised fine tuning in ASR, and (iii) random initialization + supervised fine tuning.
% In this subsection, we analyze the effects of initialization and fine-tuning through transfer learning for ASR with the proposed attention based approach. 
Table \ref{tab:unsup_vs_sup_onlyAcFB} shows the effects of acoustic FB initialization on ASR. All the features are trained with {\textcolor{black}{the}} proposed relevance weighting based model. \textcolor{black}{For the random initialization, we sample the center frequencies uniformly  at random or use a sigmoidal mapping function on the uniform random variable.} It can be observed that unsupervised initialization alone of acoustic FB parameters (and no fine tuning) does not improve over baseline. The random initialization of acoustic FB followed by supervised fine-tuning for ASR gives considerable improvement in ASR performance. \textcolor{black}{The random initialization of acoustic FB means on sigmoidal frequency scale performs better than the uniform scale with no fine tuning.} The approach of unsupervised initialization with supervised fine-tuning (approach followed in this paper) gives the best ASR performance among all choices considered here.

\subsection{\textcolor{black}{Choice of hyper-parameters }}
\textcolor{black}{We experiment with various choices of hyper-parameters in acoustic FB learning. The effect of context length of the input patch ($t$) is analyzed through ASR performance with the proposed [A-R, M-R] approach. The Aurora-4 dataset is used with different context length of $t=21$, $51$, $81$, $101$ and $131$. From the results reported in Table \ref{tab:effect_contextLen}, it can observed that the performance improves with increasing the context length, with the best performance for $t=101$ and $t=131$.}

\textcolor{black}{We also observe the effect of acoustic filter length $k$. While $k=8$ms yields a frequency resolution of $125$Hz, a $4$ms long kernel has a frequency resolution of $250$Hz and a $16$ms long kernel has a frequency resolution of $62.5$Hz. The baseline features like mel filter bank use windowing in the frequency domain with mel-spaced filters. The filters in the lower frequency range in the mel-scale have higher resolution (around $100$ Hz).  From Table \ref{tab:centerFreq_acFB_diffLen}, it can be observed that while the shorter filter length degrades the ASR performance, the increased filter length beyond $8$ms does not improve the ASR performance.}
% \textcolor{black}{In addition, the total number of parameters in the mel-scale baseline (MFB) is around $27.92\times 10^6$ and in the proposed approach is $27.98 \times 10^6$. In terms of total number of parameters involved, the proposed approach only increases the number of parameters by $0.2$\%.}

% \begin{table}[h]
%     \caption{Unsupervised learning vs. Supervised learning of representation}
%     \label{tab:unsup_vs_sup}
%     \centering
%     \begin{tabular}{c|c}
%     \hline
%         Type of learning & ASR \\\hline
%         Unsupervised Initialization (\& no fine tuning) & 10.8 \\
%         Unsupervised Initialization + supervised fine tuning & 9.9 \\
%         Random Initialization + supervised fine tuning & 9.8 \\
%         Random Initialization + No fine tuning & 11.3 \\\hline 
%     \end{tabular}
% \end{table}{}

\begin{table}[t]
    \caption{Unsupervised learning
    % (Only Ac FB) 
    vs. Supervised learning of acoustic filterbank with [A-R,M-R] configuration \textcolor{black}{on Aurora-4 dataset}.}
    % Mod. filter always rand init + sup. updated}
    \label{tab:unsup_vs_sup_onlyAcFB}
    \centering
    \begin{tabular}{c|c}
    \hline
        Type of acoustic FB learning & ASR \\\hline
         \textcolor{black}{ Random Initialization on uniform scale + No fine tuning} & \textcolor{black}{11.9} \\
         Random Initialization on sigmoidal frequency scale + No fine tuning & 11.2 \\
          
        Unsupervised Initialization (\& no fine tuning) & 10.9 \\
        Unsupervised Initialization + supervised fine tuning & \textbf{9.6} \\
        Random Initialization on sigmoidal scale + supervised fine tuning & 9.9 \\\hline
       
    \end{tabular}
\end{table}{}
\begin{table}[t]
    \centering
    \caption{\textcolor{black}{Effect of context length of the input patch (value of $t$) on Aurora-4 ASR performance with the [A-R,M-R] approach.}}
    \begin{tabular}{c|c}
    \hline
        \textcolor{black}{Context length (value of} $t$) & \textcolor{black}{ASR (WER in} \%) \\\hline
        21 & 10.2 \\
        51 & 10.1 \\
        81 & 10.0 \\
        101 & \textbf{9.6} \\
        131 & 9.6 \\\hline
    \end{tabular}
    \label{tab:effect_contextLen}
    \vspace{-0.2cm}
\end{table}
% \begin{figure}[t!]
%     \centering
%     \includegraphics[scale=0.35]{Figures/center_freq_means_diffFilterLength.eps}
%     \caption{\textcolor{black}{Learnt center frequency with different filter length (value of $k$) in acoustic FB layer with Aurora-4 dataset training.}}
%     \label{fig:centerFreq_acFB_diffLen}
% \end{figure}
\begin{table}[t]
    \centering
    \caption{\textcolor{black}{Effect of different filter length ($k$) in acoustic FB layer on ASR performance with Aurora-4 dataset.}}
    \begin{tabular}{c|c}
    \hline
    \textcolor{black}{Filter length (value of $k$)} &  \textcolor{black}{ASR (WER in \%)} \\\hline
    64  ($4$ms) & 10.6 \\
    128 ($8$ms) &\textbf{ 9.6} \\
    256 ($16$ms) & 9.7 \\
    \hline
    \end{tabular}
    \label{tab:centerFreq_acFB_diffLen}
\end{table}
\begin{figure}[t!]
    \centering
    \includegraphics[trim={0.5in 0 0 0.2in}, clip, scale=0.325]{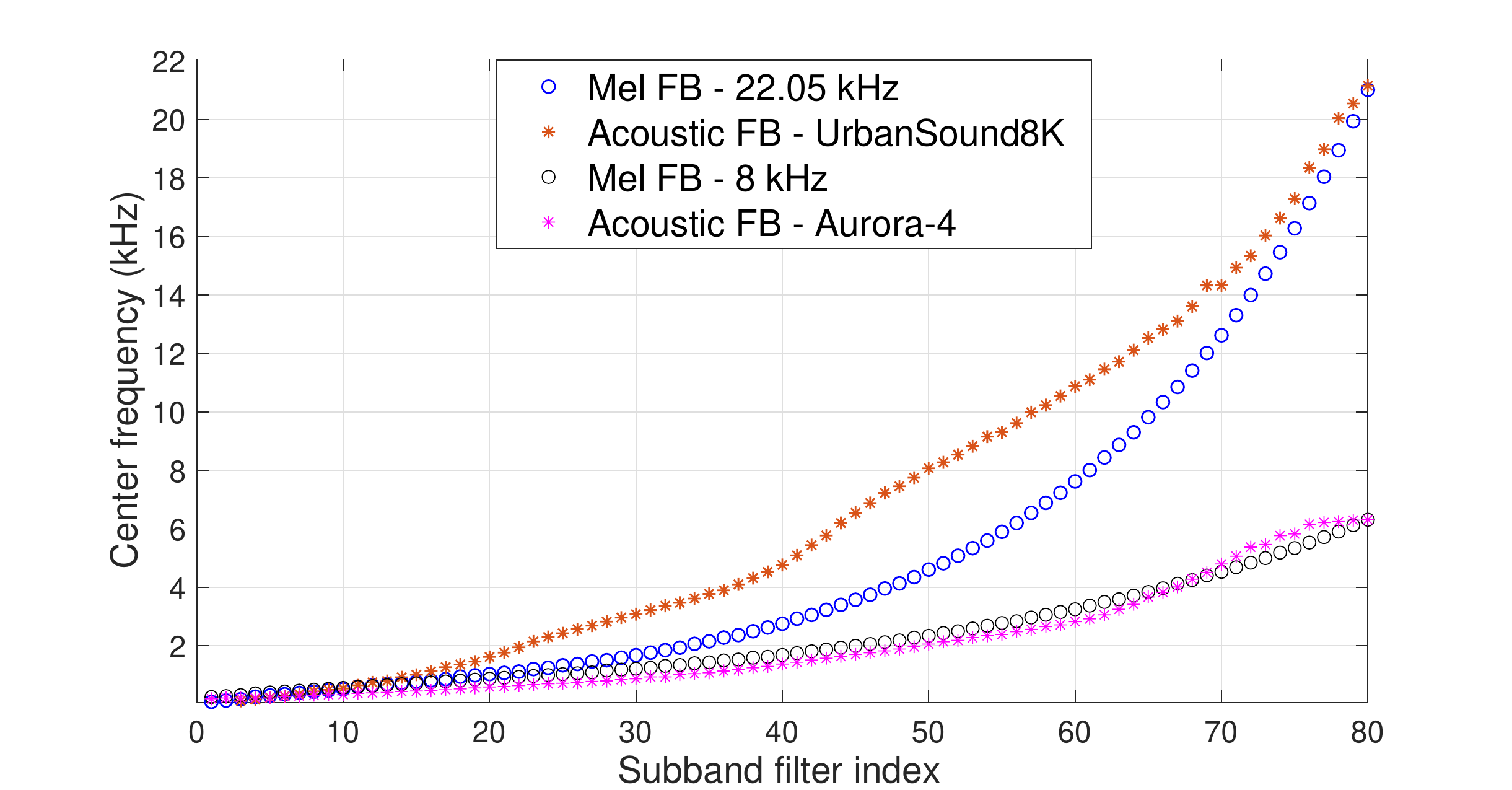}
    % \vspace{-0.1cm}
    \caption{\textcolor{black}{Comparison of center frequency of acoustic FB learned using the proposed approach with those of mel FB for audio (urban sounds) and speech datasets. }
    % The filter indices are ordered in their increasing order of center frequency.
    }
    \label{fig:center_freq_acoustic_urban}
    %  \vspace{-0.4cm}
\end{figure}

\subsection{\textcolor{black}{Urban Sound Classification}}
\textcolor{black}{The spectro-temporal characteristics of the different types of audio signals are different, and hence, the learned time-frequency representation should capture these distinctive properties for efficient sound classification. The proposed two stage approach of representation learning is explored for urban sound classification task. The same architecture used for speech (shown in Figure \ref{fig:block_diag}) is used for audio representation learning. The input to the network is a short audio snippet and the network is trained for classifying $10$ urban sound classes. Following the acoustic FB layer and the modulation filtering layer (with the two relevance weighting sub-networks), the model consists of a CNN-LSTM for the USC task (the LSTM backend provided an improved performance for the USC task but not for the ASR task).}

\begin{table}[t!]
    \centering
    \caption{\textcolor{black}{Classifier accuracy (\%) in UrbanSound8K database. Here the different sound categories are Air conditioner (AI), Car horn (CA), Children playing (CH), Dog bark (DO), Drilling (DR), Engine idling (EN), Gun shot (GU), Jack hammer (JA), Siren (SI) and Street music (ST).}
    % {\textcolor{red}{Will update numbers with 10-fold CV, trainings going on}}
    }
    \label{tab:urbanSoundClassification_results}
    \vspace{-0.15cm}
    \resizebox{\columnwidth}{!}{
   
    \begin{tabular}{c|c|c|c|c|c|c|c|c|c|c|c}
    \hline
    \multirow{2}{*}{Rep.} & \multicolumn{11}{c}{{Accuracy} (\%)}  \\ \cline{2-12}
    & AI & CA & CH & DO & DR & EN & GU & JA & SI & ST & Avg\\ \hline
    MFB,M & 32 & 61 & 50 & 65 & 45 & 44 & 54 & 44 & 66 & 61 & 52\\
    A,M &  32 & \textbf{67} & 60 & 68 & 63 & 55 & \textbf{82} & 48 & 71 & 65 & 58\\
    % A-R & 64 \\
    A-R,M-R & \textbf{40} & {62} & \textbf{60} & \textbf{68} & \textbf{66} & \textbf{56}	 & {79} & \textbf{59} & \textbf{74} & \textbf{67} & \textbf{62}\\ 
    M3\cite{dai2017very} & - & - & - & - & - & - & - & - & - & - & 58 \\\hline
    \end{tabular}
    }
    \vspace{-0.1cm}
\end{table}
\begin{figure*}[t!]
    \centering
    \vspace{-0.1cm}
    \includegraphics[trim={4cm 0 4cm 0.1cm}, clip, width=0.9\linewidth]{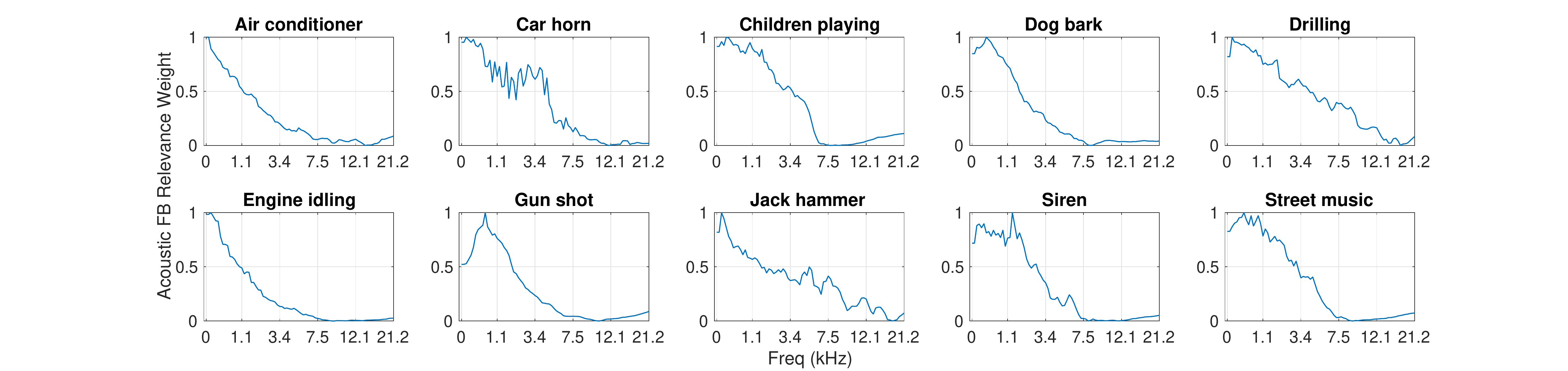}
    \vspace{-0.2cm}
    \caption{\textcolor{black}{The normalized acoustic FB relevance weight profile ($\boldsymbol{w}_a$ in Fig. \ref{fig:block_diag}) averaged over audio sounds from UrbanSound8K dataset.}}
    \label{fig:avg_acAttn_urban}
    \vspace{-0.1cm}
\end{figure*}
\begin{figure}[t]
    \centering
        \includegraphics[trim={2.3cm 0cm 1.8cm 0.6cm}, clip, width=\columnwidth]{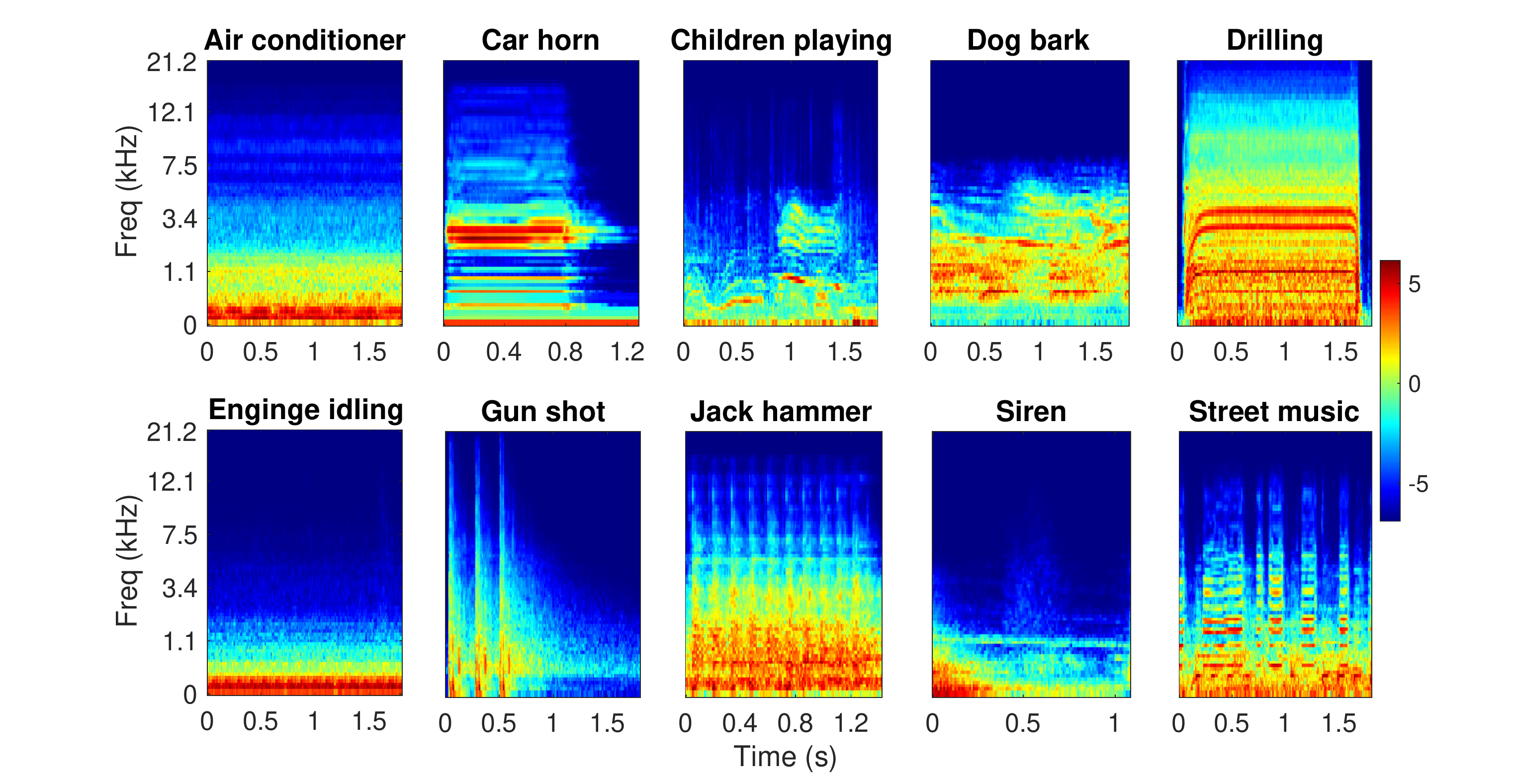}
        \vspace{-0.6cm}
        \caption{Time-frequency representation learned by the model ($\boldsymbol{x}$ in Fig. \ref{fig:block_diag}) plotted for a file from each urban sound class.
        % from UrbanSound8k dataset.
        }
        \label{fig:spec_eachClass_urbanSound}
        \vspace{-0.35cm}
\end{figure}

\subsubsection{\textcolor{black}{Dataset}}
\textcolor{black}{The USC system is trained using PyTorch \cite{paszke2017pytorch} using UrbanSound8K dataset \cite{salamon2014dataset}. It contains $8732$ sound clips sampled at $44.1$kHz, with duration up to $4$s divided into $10$ sound classes: air conditioner (AI), car horn (CA), children playing (CH), dog bark (DO), drilling (DR), engine idling (EN), gun shot (GU), jackhammer (JA), siren (SI), and street music (ST). The best trained model is chosen through cross-validation loss. In our work, we use $9$ folds for training purpose, and the results are reported with average of $10$-fold cross validation.}

\textcolor{black}{Fig. \ref{fig:center_freq_acoustic_urban} shows the center frequency ($\mu_i$ values sorted in ascending order) of the acoustic filters obtained for the USC task. A comparison with the mel filter center frequencies as well as learnt center frequency from Aurora-4 speech data is also provided for reference. The center frequency of learned filters in audio data deviate from the mel counterparts. In particular, the learned representations have more emphasis on higher frequencies compared to the mel spectrogram.}

\textcolor{black}{Table \ref{tab:urbanSoundClassification_results} reports the average USC performance for baseline system and the proposed two-stage approach of representation learning. As seen in the results, the learnt acoustic FB features (A) alone (without any relevance weighting) performs considerably better than the fixed MFB features, with an absolute improvement of $6$\%. The 2-stage (A-R,M-R) approach further gives improvement in almost all the classes with absolute improvement of $10$\% on average over baseline MFB system. In addition, the results are better than the very deep CNN model (M3) with large number of kernels ($384$) on raw audio waveform reported in \cite{dai2017very}.}

\subsubsection{\textcolor{black}{Analysis}}
\textcolor{black}{For the audio sounds in the Urbansound8K dataset, we analyze the learned time-frequency representation of each class in Fig. \ref{fig:spec_eachClass_urbanSound}.
As seen here, air conditioner, engine idling and siren sounds have  mostly low frequency content. The car horn, drilling and street music sounds have multiple peaks at different frequencies. The drilling and jack hammer sounds have higher frequency range spanning till $20$kHz.}

\textcolor{black}{The acoustic FB relevance weights for the audio signals are analyzed in Fig.~\ref{fig:avg_acAttn_urban}. It can be observed that most of the sounds exhibit higher relevance weight for low to mid acoustic frequencies, with drilling and jackhammer sounds having significant weight values till the end of the spectral range.
The air conditioner and engine idling are low frequency sounds with peak primarily in between $10-200$Hz. The car horn, siren and street music sounds exhibit multiple frequencies and hence, have higher value for relevance weights till around $4$kHz. The gun-shot sound weight profile exhibits an exclusive peak at around $550$Hz. }

\color{black}{\subsection{Contributions Beyond Prior Works}}
\textcolor{black}{
The self-attention approach \cite{lin2017structured} applies a fixed neural layer on inputs (typically hidden layer activations indexed by time) to generate a single scalar value. A softmax operation applied on the scalar values over time provides the attention weights. The attention weights are used to generate a linear combination of the embeddings. In the proposed work, while a softmax operation on scalar relevance weight values are performed, we do not perform any linear combination of the inputs. Thus, the scalar values act as a relevance weights (gain) that modulate the information flow to the subsequent layers. A higher value of the weight signifies more relevance to the corresponding input stream (sub-band in acoustic FB layer or modulation filter in the modulation FB layer). In addition, self-attention in LSTM networks are typically performed at the end of the encoding layers (closer to the target layer) or as multiple modules of self-attention in transformer networks~\cite{vaswani2017attention} which carry all the way upto the target layers. In the proposed framework, the relevance weighting is only performed in the initial layers which are closer to the input  representations. Hence, the relevance weights offer interpretability in terms of speech/audio characteristics.  }\

\textcolor{black}{
The gated neural network approach \cite{shazeer2017outrageously} explored multiple parallel ``expert'' networks followed by a gating network which performs a convex combination of the outputs from the networks. Again, similar to self-attention, the model only propagates the combined output. }

\textcolor{black}{
The ASSERT network \cite{lai2019assert} for spoof detection proposes an attention filtering layer. This approach operates on speech spectrograms of the entire utterance (typically few seconds) and generates an attentional heat-map using a deep convolutional neural network with a U-net model containing multiple levels of dialation, downsampling and upsampling layers along with skip connections. The attentional heat-map is then element wise multiplied on the input utterance spectrogram and added with input spectrogram. The entire model is trained for spoof detection task. In the proposed approach, the model does not generate a heat-map on all components. It rather provides a scalar relevance weight for each sub-band (acoustic FB layer) or for each modulation filter (modulation FB layer). In addition, the proposed model does not add the weighted input again with the input but rather propagates the weighted feature stream directly to the subsequent layers. Further, the authors in ASSERT use the attention filtering only on the input log-magnitude spectrogram. The ASSERT attention filtering will be more cumbersome than the proposed approach when incorporated in modulation filtering layers as the attentional heat-map needed would be 3-D in nature.}

\vspace{0.5cm}
\section{Summary}{\label{sec:summary}}
In this paper, we have proposed a framework for relevance weighting based representation learning from raw waveform.

\subsection{Key Contributions}
% The details of the proposed model and its application to ASR has been presented. With several speech recognition experiments in a multi-condition training setup, we have also illustrated the performance benefits of the proposed approach compared to baseline  methods as well as several other noise robust front-ends proposed in the past. The analysis of acoustic and modulation relevance weights for different phonemes is crucial contribution in terms of interpretability of the representation learning.
The key contributions of the work can be summarized as follows:
\begin{itemize}
    \item Posing the representation learning problem in an interpretable filterbank learning framework using relevance weighting mechanism from raw waveform.
    \item Using a two-step process for learning representations: first step to obtain time-frequency (spectrogram) representation from raw waveform; second step to perform modulation filtering on first step output. Both steps have separate relevance weighting sub-networks, which are differentiable and can be easily integrated in the backpropagation learning.
    \item Implementing the acoustic filterbank in the first convolutional layer of the proposed model using a parametric cosine-modulated Gaussian filter bank whose parameters are learned. A relevance sub-network provides relevance weights to acoustic FB.
    \item The second convolutional layer performs modulation filtering over the first layer output. Another relevance sub-network provides relevance weights to the modulation feature maps.
    \item ASR experiments on a variety of speech datasets containing noise and reverberation showing the benefits of the proposed approach. 
    % \item Illustrating the benefits of the proposed approach being consistent in semi-supervised ASR training as well.
    \item Analysis of the relevance weights reveal that the weights capture underlying phonetic content in speech.
    \item \textcolor{black}{The proposed 2-stage framework is extended to urban sound classification task where it shows improved performance. The weight analysis reveals that distinct audio characteristics  are captured by the relevance weights}. \footnote{The implementation of the proposed approach is available at https://github.com/PurviAgrawal/Interpretable\_rawWaveform\_relevanceWeighted}
    % \item Exploring the presence of common speech representations which can be learned from any one of the corpus and generalized to other datasets.
    % \item Analysing the adaptive nature of the proposed relevance weighting towards unseen noise conditions in test.
\end{itemize}

{\color{black}{\subsection{Limitations} Some limitations of the proposed work include, }
\color{black}{
\begin{itemize}

    \item \textcolor{black}{ The proposed approach is empirical. The model is proposed and analyzed based on performance metrics.}
    
    \item \textcolor{black}{The total number of parameters in the mel-scale baseline is around $27.92$M while in the proposed approach the number of trainable parameters is $27.98$M. In terms of total number of parameters involved, the proposed approach only increases the number of parameters by $0.2$\%. In terms of computational time, we performed an experiment with $100$ test files from the Aurora-4 corpus and measured the computation time for forward pass through the model for the baseline system as well as the proposed relevance weighted model. The computation time for the proposed model increased by $20$\% ($171$sec. of computational time for the proposed method versus $136$sec. for the baseline mel-spec system).}
\end{itemize}}
% This evaluation was performed on an Intel CPU with single core. 
% On a GPU system, the computational times showed lesser of an increase for the proposed approach. 
\vspace{-0.4cm}
\textcolor{black}{\subsection{Future Work}} 
\textcolor{black}{
The relevance weighting should also incorporate some level of feedback if the proposed model is  to imitate biology in true sense. In addition, other gross statistics of the signal maybe potentially useful in deriving the relevance weights.  Furthermore, extending the approach to representation learning for other machine learning tasks like speaker recognition and language modeling are part of future considerations.
}} \\
% \color{black}{
% \section*{Acknowledgment}

% This work was partly funded by grants from the Ministry of Human Resource and Development (MHRD), Government of India, and the Department of Atomic Energy (DAE) project (DAE/34/20/12/2018-BRNS/34088).
% }
% \begin{center}
%     \begin{figure}[t]
%         \centering
%         \includegraphics[trim={0 0 0 1.6cm}, clip, scale=0.34]{Figures/bar_plot_semi_sup.pdf}
%         % \vspace{-0.1cm}
%         \caption{ASR performance in WER (\%) for Aurora-4 database (avg. of 14 test conditions) using lesser amount of labeled training data.}
%         \label{fig:semisup}
%         % \vspace{-0.2cm}
%     \end{figure}
% \end{center}
% \section*{Acknowledgment}

% This work was partly funded by grants from the Ministry of Human Resource and Development (MHRD), Government of India, and the Department of Atomic Energy (DAE) project (DAEO0205).

\bibliographystyle{IEEEtran}
\bibliography{refer}

\vspace{-0.3in}
%  \begin{IEEEbiography}[{\includegraphics[width=1in,height=1.25in,clip,keepaspectratio]{purvi_pic_taslp.jpg}}]{Purvi Agrawal}
  \begin{IEEEbiography}[{\includegraphics[width=1in,height=1.25in,clip,keepaspectratio]{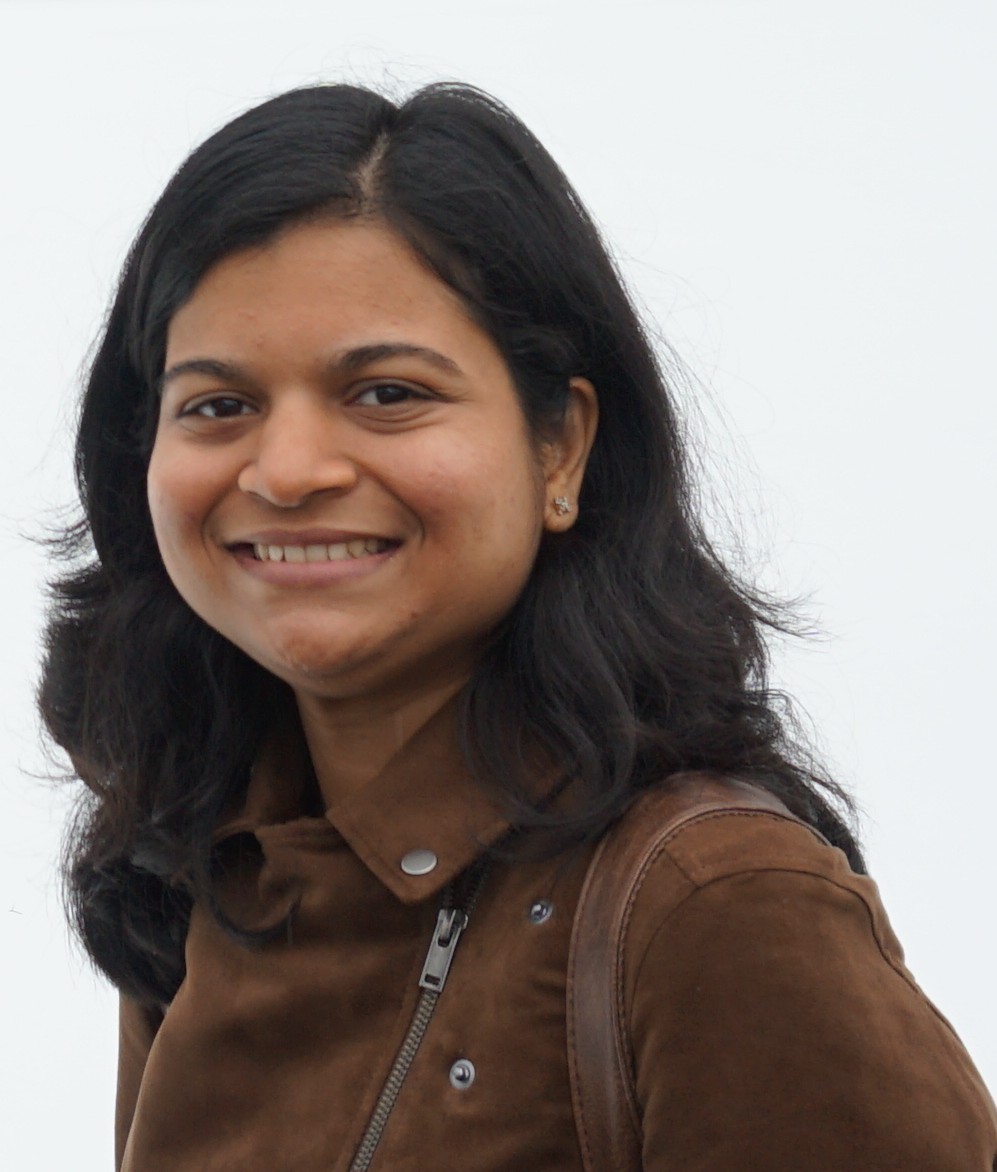}}]{Purvi Agrawal}
 is a PhD scholar at Learning and Extraction of Acoustic Patterns (LEAP) lab, Electrical Engineering, Indian Institute of Science (IISc), Bangalore. Prior to joining IISc, she obtained her Master of Technology in Communications from DA-IICT, Gandhinagar in 2015. She obtained her Bachelor of Technology in Electronics and Communication from College of Technology and Engineering, Udaipur in 2013. She has also worked in Sony R \& D Labs, Tokyo in 2017. Her research interests include machine learning for signal processing, unsupervised learning and applications, and robust speech recognition. She is a student member of IEEE.
 \end{IEEEbiography}
\vspace{-0.5in}
\begin{IEEEbiography}[{\includegraphics[width=1in,height=1.25in,clip,keepaspectratio]{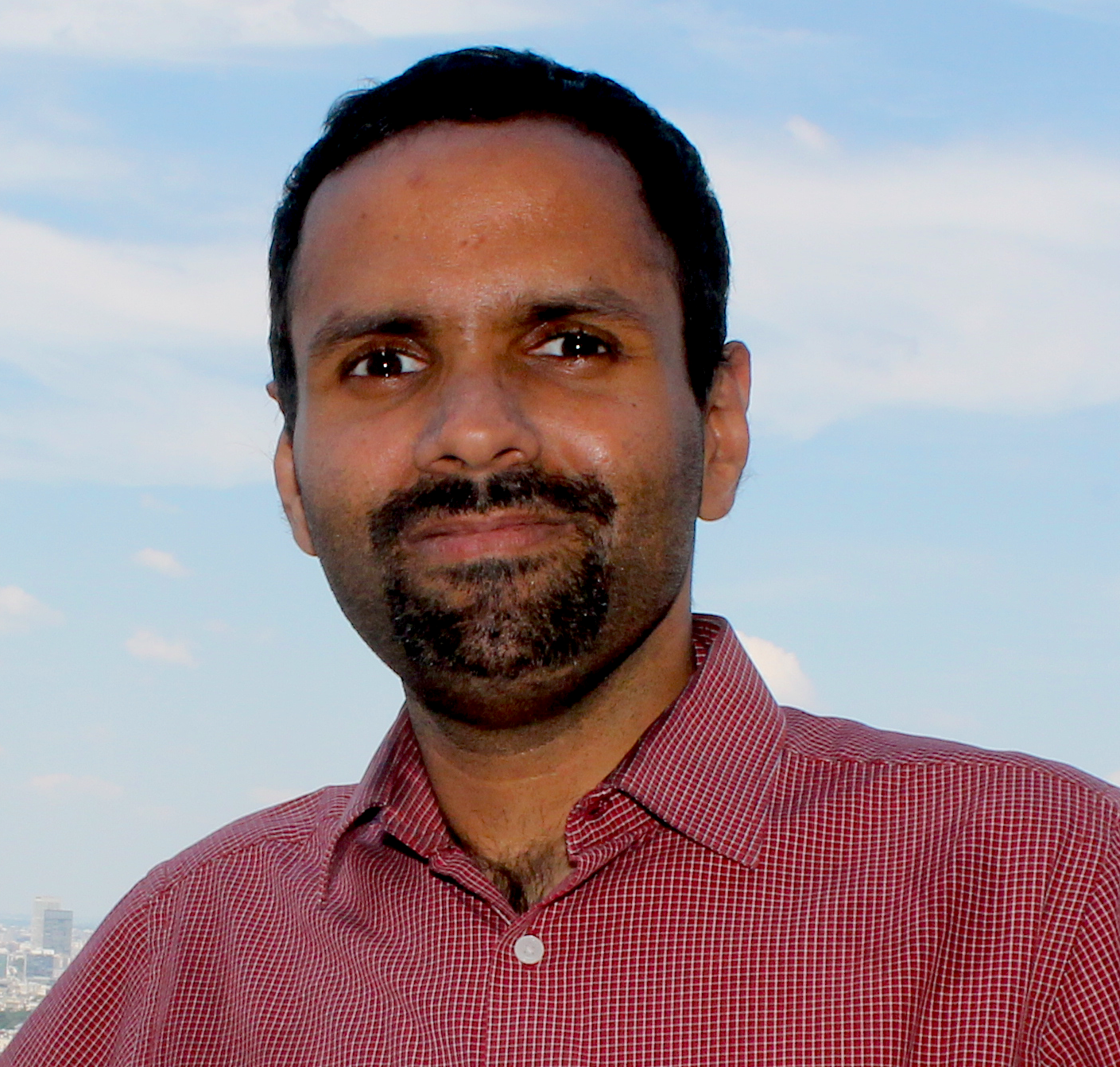}}]{Sriram Ganapathy}
is a faculty member at the Electrical Engineering, Indian Institute of Science, Bangalore, where he heads the activities of the learning and extraction of acoustic patterns (LEAP) lab. Prior to joining the Indian Institute of Science, he was a research staff member at the IBM Watson Research Center, Yorktown Heights. He received his Doctor of Philosophy from the Center for Language and Speech Processing, Johns Hopkins University. He obtained his Bachelor of Technology from College of Engineering, Trivandrum, India and Master of Engineering from the Indian Institute of Science, Bangalore. He has also worked as a Research Assistant in Idiap Research Institute, Switzerland from 2006 to 2008.  At the LEAP lab,  his research interests include signal processing, machine learning methodologies for speech and speaker recognition and auditory neuro-science. He is a subject editor for the Speech Communications journal and a senior member of the IEEE. 
\end{IEEEbiography}

% \begin{IEEEbiographynophoto}{Sriram Ganapathy}

\end{document}